\newif\ifShowKeys
\ShowKeystrue
\newif\ifshowtikz
\showtikztrue
\showtikzfalse   

\documentclass[11pt]{article}
	\pdfoutput=1
\usepackage[T1]{fontenc}

\usepackage{listings}
\lstnewenvironment{arkady}  
{\lstset{language=C,frame=trbl,basicstyle = \footnotesize \ttfamily , breaklines = true,showstringspaces=false}}{}

\usepackage{datetime}
\usepackage{comment}					

\usepackage[usenames,dvipsnames]{xcolor}
\usepackage[setpagesize=false,pagebackref=false, 
linktocpage, bookmarksopen=true, colorlinks=true, 
linkcolor=Maroon,citecolor=Maroon,urlcolor=Maroon]{hyperref}

\usepackage[parsep]{collref}				

\usepackage{amsmath, amssymb,amsthm}
\usepackage{stackrel}
\numberwithin{equation}{section}
\usepackage{bm,environ,mathrsfs,array,arydshln}
\usepackage{booktabs,float,slashed}
\usepackage{appendix}
\usepackage[mathcal]{euscript}
\usepackage{tensor} 						
\usepackage{mathabx}
\usepackage[vcentermath]{youngtab}
\usepackage{simpler-wick}

\usepackage{graphicx,epsfig,epic}
\usepackage{tikz}
\usepackage{tikz-feynman} 

\allowdisplaybreaks

\usepackage{framed}						
\definecolor{shadecolor}{rgb}{0.9996078, 0.984314, 0.960784}
\definecolor{framecolor}{rgb}{0,0,0}
\definecolor{TFTitleColor}{RGB}{1,1,1}

\newenvironment{frshaded}{%
    \MakeFramed {\FrameRestore}}%
    {\endMakeFramed}

\definecolor{myred}{RGB}{233, 33, 45}

\newenvironment{remark}
  { \begin{list}{}%
      {\setlength{\leftmargin}{0.5cm}%
       \setlength{\rightmargin}{0.5cm}}%
   \item\relax\(\triangledown\)\textbf{ Remark }}
  {\end{list}}

\newcommand{\bs}{\begin{frshaded}}			
\newcommand{\es}{\end{frshaded}\noindent}

\def\ba#1\ea{\begin{align}#1\end{align}}		        
\newcommand{\be}{\begin{equation}}
\newcommand{\ee}{\end{equation}}
\newcommand{\bea}{\begin{equation} \begin{aligned}} 
\newcommand{\eea}{\end{aligned} \end{equation}}
\newcommand{\mc}{\mathcal }

\newcommand{\wt}{\widetilde}

\newcommand{\mk}{\mathfrak}
\newcommand{\la}{\label}
\newcommand{\eps}{\varepsilon}

\newcommand{\lp}{\notag \\ & }

\DeclareMathOperator{\tr}{\text{tr}}

\newcommand{\N}{\mathcal N}

\newcommand{\sql}{\sqrt\l}
\renewcommand{\l}{\lambda}
\newcommand{\gym}{g_{\scalebox{0.45}{\text{YM}}}}

\newcommand{\adstwo}{AdS$_{2}$ }

\newcommand{\LDR}{\ell}
\newcommand{\sss}{{\rm S}}
\newcommand{\gxx}{{\rm G}}
\newcommand{\G}{\Gamma}
\newcommand{\w}{{\rm w}}

\def \np {\newpage}
\def \ed {\np \small
\baselineskip 11pt
\bibliography{BT-Biblio}
\small
\bibliographystyle{JHEP-v2.9}
\end{document}
}
\def \iffa  {\iffalse}
\def \te {\textstyle}
\newcommand{\rf}[1]{(\ref{#1})}
\def\ov{\over}
\def \ci {\cite}
\def \foot {\footnote}
\def \bi{\bibitem}
\def\la{\label}\def \a {\alpha}
\def\foot{\footnote}
\def \adss {AdS$_5 \times S^5$\ }

\def \s {\sigma} \def \del {\partial} 
\def \ha {\tfrac{1}{2}}
\def \DD {{\rm D}}

\def \OO {{\cal O}}

\def \no {\nonumber}
  
 \def \g {\gamma}
 
\newcommand{\Pf}{\mathcal{P}}

\newcommand{\h}{\mathrm{h}}

\begin{document}
\begin{titlepage}
\begin{tabbing}
\hspace*{10.5cm} \=  \kill 
\>  Imperial--TP--2025--SK--02 \\
\> 
\end{tabbing}

\vspace*{15mm}
\begin{center}
{\Large\sc  \bf     
Strong coupling expansion of  $1\ov 2$  BPS Wilson  loop  in SYM theory \vskip 3pt
   and   2-loop   Green-Schwarz  string   in AdS$_5 \times S^5$  }
\vspace*{10mm}

M. Beccaria$^{a}$, \ \ S.A.  Kurlyand$^{b}$,\ \ \ A.A. Tseytlin$^{b,}$\footnote{Also at  ITMP and Lebedev Inst.}

\vspace*{4mm}
{\small
	
${}^a$ Universit\`a del Salento, Dipartimento di Matematica e Fisica \textit{Ennio De Giorgi},\\ 
		and INFN - sezione di Lecce, Via Arnesano, I-73100 Lecce, Italy
			\vskip 0.3cm
${}^b$ Abdus Salam Centre for Theoretical Physics,\\ Imperial College London,  SW7 2AZ, U.K.
			\vskip 0.01cm
\vskip 0.2cm {\small E-mail: \texttt{matteo.beccaria@le.infn.it,\ s.kurlyand23@ic.ac.uk, \ tseytlin@ic.ac.uk}}
}
\vspace*{1.8cm}
\end{center}
\begin{abstract}  
	{The exact localization result for the expectation  value of the $1\over 2$  BPS circular  Wilson  loop  in ${\cal N}=4$  SYM  theory is  given  in the planar limit by the  famous Bessel function  expression: $\langle W\rangle = {2N\ov \sqrt \lambda } I_1 ( \sqrt \lambda)$. Expanded  in large $\lambda$ and expressed in terms of the  AdS$_5 \times S^5$  string  tension $T= {\sqrt \lambda \over 2\pi}$ this gives  $\langle W\rangle = {\sqrt T\over 2\pi g_s}  e^{2\pi T} (1- {3\over 16 \pi} T^{-1} + ...)$.The  exponential is matched by the  value of the action  of the string with the  AdS$_2$  world volume  while   the prefactor comes from the 1-loop   GS string correction. Here we  address the question 
 of how  the subleading $T^{-1}$ term  could  be  reproduced by the 2-loop  correction in the  corresponding partition function of the AdS$_5 \times S^5$ GS string expanded near the AdS$_2$   minimal surface. We  find that the string  correction contains  a non-zero UV  logarithmic divergence  implying that comparison with   the  SYM  result  requires a particular subtraction prescription. We discuss  implications of this conclusion for checking the AdS/CFT duality at strong coupling. }
   \end{abstract}
\vskip 3.5cm
\end{titlepage}
{\small
\makeatletter
\newcommand*{\toccontents}{\@starttoc{toc}}
\makeatother
\toccontents
}
\def \adst {AdS$_2$ }
\def \eptt {\bar \eps}
\def \ept  {\eps'}
\def \adstwo {\adst}
  \def \rmb {{_{\rm b}}}   \def \rmf {{_{\rm f}}}  \def \rmbf {{_{\rm bf}}}
 \def \DD {{\mk D}}  \def \b {\beta } 
\def \m {\mu} \def \n {\nu} \def \L {\mc L}
 \def \ep {\eps}
\def \rd {{\rm d}}
 \def \raa  {{\rm a}} 
\def \G  {\Gamma} 
\def \hGamma {\hat \Gamma}
\def \hata {{\hat \a}}
\def \ss {{\rm s}} \def \hta {{\hat \a}} \def \htb {{\hat \b}}
\def \halpha {{\hat \alpha}}  \def \hbeta {{\hat \beta}}
\def \mC  {C}
\def \ss {{\rm s}} \def \hta {{\hat \a}} \def \htb {{\hat \b}}
\def \halpha {{\hat \alpha}}  \def \hbeta {{\hat \beta}}
\def \hG {\hat \G}
\def \ww {{\rm w}}
\def \r {\rho}
  \def \nf {N_{{\theta}}}
\def \bi {\bar I}
\def \nt {N_\theta} 
\def \rfe {{\rm G}}  \def \rh {\hat \rfe} \def \rk {\tilde  \rfe}
\def \ri {{\rm I\, }}
\def \t {\theta} 
\def \Ge {\gamma_{\rm E}} \def \rG {{\rm G}}
\def \CP {{\rm CP}}
\def \ZZ {{\mathbb Z}}


\setcounter{footnote}{0}
\section{Introduction}

Even in the most  symmetric   example of AdS/CFT  duality -- between $\N=4$ SYM and \adss  superstring -- 
the two sides   are  not on an equal footing. While the $\N=4$ SYM  has a well-defined perturbation theory, 
the \adss   superstring    in   the   Green-Schwarz  description  
is  represented  by a formally non-renormalizable 2d action  and  a priori   is not  unambiguously  defined at the 
quantum level beyond the semiclassical 1-loop approximation. 

Indeed, already in the   case of the flat target space  the  2-loop world-sheet  S-matrix   of the  GS 
string expanded near the long-string vacuum  
 contains  UV poles.  One is  thus  required to add  some  specific (divergent and finite)  counterterms order by order in loop expansion 
 in order  to   maintain its consistency 
  with  quantum  integrability \ci{Beccaria:2025xry}.\foot{Some  alternative approaches to 
  definition of string theory in \adss  are   discussed in \ci{Cho:2025coy} and refs. there.}
  
  Postulating that quantum \adss GS  theory     should have a formulation consistent   with integrability 
  implies that its classical action  should be  expected  to be supplemented  with  a particular set 
  of $2$-loop, etc.,  higher-derivative counterterms.
  While the  GS   action itself  (or the string tension) 
   should  not be renormalized    due to  symmetries of the action \ci{Metsaev:1998it}, the  logarithmic divergences do 
    appear  in the 2-loop  diagrams \ci{Roiban:2007jf}. 
    
     One    might     expect  that  in some particular    cases the GS  string partition function may still be UV finite. 
     Indeed, that  was found  to happen in the case of the 2-loop   correction to the  null cusp anomalous  dimension 
     \ci{Roiban:2007dq,Giombi:2009gd,Giombi:2010fa}. 
      This  example, however,  is  special  in 
     having flat induced 2d metric (implying that   possible covariant counterterms   may vanish). 
     As we shall   find  below,  the  2-loop  UV  poles do not cancel in the case of the  expansion near  a minimal  surface with 
     curved  \adst 
   metric.
   
   This   raises the  general question  of how  strong-coupling predictions  of SYM theory
    (coming  from localization or assumption of quantum integrability)  can be directly compared to 
    perturbative results on the \adss    string side, i.e.   how  to provide 
    independent   checks of AdS/CFT  at strong coupling  beyond the  semiclassical  1-loop   approximation. 
    There is no  known direct  way of implementing quantum integrability in the string perturbation theory 
    that would  fix  in general  the required  2-loop  and higher   counterterms allowing  to get finite 
    string  predictions  that can be compared to the SYM  side. 
    One may then need   to compute   {several }
    observables  in order   to  ``measure''  the   coefficients  of the counterterms  before  being 
    able  to make  unambiguous (scheme-independent)   predictions that  would check AdS/CFT at strong coupling.\foot{Even  ignoring the 
     issue of UV divergences on the string side,   to   provide a scheme-independent   comparison 
     of the  SYM and string  predictions one   should   compare  at  least two observables, e.g.    dimensions
      of  two  operators as  functions of the 't Hooft coupling  $\l$  in the planar limit,   and 
      eliminate $\l$   expressing   one  dimension as a function of the  other. That function can then be  matched  to the 
       one relating  the AdS energies  of  one  corresponding  string states.
       Also, in some cases,   to eliminate  possible ambiguities  in the definition of string path integral 
          it may be useful  to consider ratios of 
       partition functions, see, e.g., \ci{Kruczenski:2008zk,Forini:2015bgo,Faraggi:2016ekd,Forini:2017whz,Cagnazzo:2017sny,
Medina-Rincon:2018wjs}.} 
    
    \iffa  Compare this to quantum gravity where we
know that at two loops we need to add RRR counter term (cubic in Weyl
tensor) with a  particular  coefficient. How do we know which is the
finite part of this coefficient? We actually don't know because we
can't make comparison to something. It's very hard to imagine that at
some point there will be some observation which will fix this
coefficient. But once we   do this  (if we could do such measurement  e.g.  near black hole  and such
and fix this coefficient), we would  have predictive power -- to
explain some other QG phenomena to this order in expansion. This gives
predictive power -- standard effective field theory logic -- we can
explain other phenomena from knowing result of a  particular
measurement (that one we cannot predict from 1st principles).
\fi

In this paper we will focus on the  case of the $\ha$   BPS circular Wilson loop 
observable   in the planar limit  represented   by the \adss  string partition function on the disk 
expanded near the \adst minimal surface. We will find that the  2-loop  UV divergences do not cancel 
automatically  and thus   matching  the  known  strong-coupling result on the SYM side 
requires a particular prescription of how to  subtract the UV poles and  fix the finite part. 
It would   be interesting to see  if there are    other observables  (like particular derivatives of 
latitude WL, etc.)   that may happen to be UV finite and thus  can be directly compared to the SYM 
side.


\subsection{The setup}

One of the  postulates of the  AdS/CFT correspondence is 
the   equivalence between the expectation value $\langle W\rangle$ of 
 a  Wilson loop in $\mc N= 4$ SYM 
and the  partition function  of  the quantum superstring in  AdS$_{5}\times S^{5}$ 
defined   with   appropriate boundary conditions  \ci{Maldacena:1998im,Rey:1998ik,Berenstein:1998ij,Drukker:1999zq,Drukker:2000ep}. 
The   simplest example is provided by the $1\ov 2$  BPS circular  Wilson  loop  with  known 
 exact  gauge theory  expression \ci{Erickson:2000af,Drukker:2000rr,Pestun:2007rz} 
which in the  planar limit    is given by 
\be \la{1}
\langle W\rangle = {2N\ov \sql} I_1 ( \sql) \ . 
\ee
Expanded  in the limit of  large 't Hooft   coupling  $\l= N \gym^2$ 
	 and expressed in terms of the    \adss  string  tension $T={L^2\ov 2\pi \a'} = {\sql \ov 2\pi}$  and string coupling $g_s ={\gym^2\ov 4 \pi} $  
\rf{1}    may be written as \cite{Giombi:2020mhz}
\be \la{2} 
\langle W\rangle =  \sqrt\frac{2}{\pi}\frac{N}{\l^{3/4}}e^{\sql} 
   \Big[1-\frac{3}{8\,\sql}+...   \Big]
   = c_1 {1\ov  g_s }  \sqrt {T \ov 2\pi} e^{2\pi T} \Big[ 1- {3\ov 16 \pi}{ 1\ov   T}   +... \Big] \ , \qquad \ \ \ c_1 = {1\ov \sqrt{ 2 \pi}} \ .  \ee
 This should   represent the  large  string tension  expansion of the disk partition function 
 of the \adss superstring    near the \adst minimal surface  that ends on a circle  at the boundary of AdS$_5$
   \ci{Berenstein:1998ij,Drukker:1999zq,Drukker:2000ep}. 
   
    Indeed, 
 the  term  in the  exponential in \rf{2}  is the same as the  classical 
   string   action   evaluated on the    \adst\   induced metric with the regularized volume $V=-2\pi$. The 
    prefactor $c_1$   represents  the 1-loop  correction   given by a product of  quadratic fluctuation determinants 
     \ci{Drukker:2000ep}
   (see  also \ci{Kruczenski:2008zk,Kristjansen:2012nz,Buchbinder:2014nia}). 
   The origin of the $ \sqrt T$   factor was explained in \cite{Giombi:2020mhz}. 
   The  computation     of  the  remaining ${1\ov \sqrt{ 2 \pi}} $ factor  in \rf{2}   remains an open problem 
   as it depends on  a  precise  normalization of the GS   superstring measure  (it 
   was indirectly confirmed   by   matching   the ratio
  of   the  $\ha$   and $1\ov 4$   BPS  WL expectation values  \ci{ Medina-Rincon:2018wjs}).   
   
   Our focus here will be on the  subleading $T^{-1}$ term in \rf{2}    that should   correspond to  the  2-loop 
   contribution in the GS superstring path integral expanded near the \adst minimal surface. 
   Separating the  prefactor $Z_1 =  {1\ov  g_s }  \sqrt {T \ov 2\pi} $   in \rf{2} we may write it as 
   \ba \la{3}
  & \qquad \qquad \qquad  \langle W\rangle = Z_1 \,  e^{-F} \ , \ \ \ \ \ \ \ \ \ \   F=  - f(T) V \ , \ \ \qquad\\
  &\la{4}  f(T) \equiv  f_0 T + f_1 +  f_2  T^{-1} + ... =  - T 
   -{1\ov 2\pi}  \log c_1   +  \frac{3}{32\pi^{2}} {T}^{-1} +... \ , \qquad \ \ \ 
  f_2= \frac{3}{32\pi^{2}} \ .   \ea
   Viewing  \rf{3}  as the string partition function 
     we used that  since \adst is a homogeneous space   the corresponding free energy should be proportional to its IR-renormalized 
      volume $V= - 2\pi$  in the case of the circular boundary ($V=0$  in the case of the line). In general, 
   $V_{\rm circle}= { 1\ov z} - 2\pi$ and $V_{\rm line} =   { 1\ov z}$  where $z\to 0$ is an IR cutoff in AdS$_2$. 
   
   To avoid the step of renormalizing  the IR divergence, i.e. to 
    get a manifestly  IR  finite   quantity  one may consider the ratio\foot{Let us note that  like the \adss radius, 
   the shape and  radius of 
   the minimal surface \adst metric
     can not be renormalized   due to underlying symmetry of the theory. Thus \adst  should remain the solution of the quantum-corrected
     2d effective action also  the  2-loop and higher level.}
     $${  \langle W_{\rm circle} \rangle\ov \langle W_{\rm line}\rangle} = Z_1 \,  e^{-2\pi f(T)} \ \ . $$
   On the SYM side the expression for the circular WL  in \rf{1} 
     corresponds to the normalization choice  for which $\langle W_{\rm line}\rangle =1$.

    The \adss GS action   \cite{Metsaev:1998it} is non-linear  and  while its 1-loop  semiclassical  quantization 
    is  straightforward (see, e.g., \cite{Kallosh:1998ji,Forste:1999qn,Drukker:2000ep,Frolov:2002av}) 
    there are issues related to potential UV divergences and scheme dependence 
     at higher loop orders.
     The \adss  GS action itself  should not be renormalized 
     being protected by a high amount of symmetry  and the presence of the fermionic  WZ term   \cite{Metsaev:1998it},
     and  the 1-loop finiteness of its on-shell effective action can be checked directly \ci{Drukker:2000ep},  
        there are potential  higher loop  divergences   that  may require introduction of 
       higher-derivative  counterterms with coefficients that   need to be fixed  in a specific way
       to be consistent   with  symmetries of the model.\foot{This is  similar to the case of $T\bar T$ deformation:    to  ensure 
      that the   classical      integrability  of a  theory   extends also  to the quantum level
      one is  to add specific counterterms at each order in the  inverse tension expansion  (see a discussion in  \ci{Beccaria:2025xry}).}     
      Currently,  there is  
      no  known   way to directly  fix   such possible counterterms.\foot{If one  imagines deriving GS action 
      as a collective  coordinate action for a fundamental string   soliton in 10d supergravity  from the supergravity action one
      may   also get  some specific higher derivative corrections suppressed     by the  inverse string tension. 
      However, the supergravity  action  itself  gets   corrected
        by string $\a'$ corrections and which should  be the resulting ``bootstrapped''
     GS + corrections  action is  unclear.}

           It  could  still be  that   for some special  observables 
      the underlying symmetry of the theory  may lead  to  a  UV   finite  result  thus   allowing a  direct  comparison to 
        the finite  strong-coupling   expansion on  the dual  gauge theory  side. 
        Indeed, examples of    consistent 
              2-loop   computations in \adss  GS theory  matching the expected gauge theory  results appeared 
      in  \ci{Klose:2007rz,Roiban:2007dq,Giombi:2009gd,Giombi:2010fa}.     
   The   challenge is to find  a regularization scheme  consistent with symmetries of the theory. 
   Ideally,  one would like to use  a version of  dimensional regularization to   avoid dealing with power divergences 
   and ambiguities in choice of local  measure and  local field redefinitions (and, in particular,  gauge dependence). 
   Even  though the   GS fermions are originally   2d scalars, 
   a direct use of dimensional regularization  is problematic 
   as the GS  action involves $\eps^{\mu\nu}$ symbol  in the fermionic WZ  term. 
   
   Once the GS action   is expanded near a particular  bosonic background, the 
   fermion  kinetic  term takes a 2d Dirac form  so that  the fermions  may be interpreted  as a collection of 2d spinors. Then 
   the use of dimensional regularization 
    would  not be  consistent with the effective   2d   global  \adst  supersymmetry \ci{Drukker:2000ep}
  of the  gauge-fixed   world sheet theory.
   One may  still  try   to use some   version of  dimensional reduction 
  regularization \cite{Siegel:1979wq} 
   like it was done in  \ci{Roiban:2007dq,Giombi:2009gd,Giombi:2010fa}.\foot{The examples   considered in  \ci{Roiban:2007jf,Roiban:2007dq,Giombi:2009gd,Giombi:2010fa,Bianchi:2014ada} were   different from the present one  in 
  having  flat induced metric   and thus flat-space  propagators for 2d fields 
   (but  lacking 2d  Lorentz symmetry in the interaction terms).
   The  special  Catalan-constant contribution
  (matching gauge-theory integrability prediction  \cite{Basso:2007wd}) was coming from  non-trivial 2-loop  integrals that could 
  be  isolated from  UV  divergent and scheme-dependent parts. In fact, 
  the logarithmic UV divergences were  found  to cancel in  the gauges used in 
 \ci{Roiban:2007dq,Giombi:2009gd,Giombi:2010fa}. This will  no  longer  be 
  so   for  the circular WL  case   discussed below  where all terms   will have rational coefficients 
and thus  will be   on an equal footing.}  
   In  general,    having   curved \adst  background   metric  requires special  care in  choice of regularization prescription 
    even in simpler  2d  QFT models   like the super-Liouville one \ci{DHoker:1983xdu,Beccaria:2019dju} 
    (see also \ci{Buchbinder:2013nta,Beccaria:2019stp,Beccaria:2019mev,Beccaria:2020qtk}). 
    The ambiguity in  choice  of quantization  is  to be fixed by extra  constraints 
    like preservation of hidden symmetries of the
    path integral or  expected   symmetries of boundary correlators.

  \subsection{Results}
  
   Fixing the static   gauge on bosons  and  the  corresponding  ``adapted''  $\kappa$-symmetry gauge on the fermions 
   (generalizing the flat-space  gauge   used, e.g., in \ci{Beccaria:2025xry})  one   finds 
   that the  Lagrangian  for the  physical  8+8  GS string fluctuations 
    takes  the following symbolic form \ci{Drukker:2000ep,Giombi:2017cqn}\foot{We assume that the radius of \adst  is 1 and rescale
    the  2d fields by  square root of the tension  $T$.}
   \ba  
   L = & (\del x^i)^2  + m^2_{\rmb}  x^2 + (\del y^a)^2  + i \bar \theta ( \slashed \nabla + m_{\rmf}) \theta  + T^{-1} \Big[ (\del x)^4  + x^2 (\del x)^2 + x^4  + (\del x)^2 (\del y)^2  \no  \\ 
 &  +   (\del y)^4 + y^2 (\del y)^2 + (\del x  \del x  + x^2) (\theta \nabla \theta + \theta^2 ) + ...
     + \theta  \theta \theta  \nabla \theta   +   \theta\nabla  \theta\theta  \nabla \theta + ...     \Big]  + \OO ( T^{-2}) \ . \la{5}
   \ea
  Here all indices   are contracted with the induced 
    \adst metric and derivatives are  2d covariant so that  there is a manifest \adst   symmetry 
  with fermions $\theta$  treated as a set of 8  Majorana  2d fermions. 
   $x^i$ represent 3  transverse fluctuations in  AdS$_5$   with mass $m^2_\rmb  =2$,   $y^a$   are massless 
   fluctuations in $S^5$ 
    and   $\theta$   have  mass $m^2_\rmf =1$. 
    
    The important simplification (compared to the examples in  \ci{Roiban:2007jf,Roiban:2007dq,Giombi:2009gd,Giombi:2010fa}) 
     is that  in  the static $\kappa$-symmetry gauge   we will use  below 
    there are no cubic interaction terms in \rf{5} 
     and thus  all 2-loop  graphs  contributing to  the free energy  will be just the 
      ``double-bubble''  ones   
%
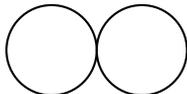
\be
\nonumber
\begin{tikzpicture}[thick,baseline=0,scale=0.8]
	\draw (0,0) circle(0.75);
	\draw (2*0.75,0) circle(0.75);
\end{tikzpicture}
\ee 
     with each of the two  propagators  being either bosonic or fermionic. 
     As the \adst  propagators  (and their derivatives) 
      taken at the coinciding points are constant,   the remaining 
        integral over the bulk point   is trivial, giving just  a factor of the \adst volume $V$ as in \rf{3}.    
    

We will use  the  dimensional  reduction regularization, i.e. will  keep the
  field indices and thus the   tensor and spinor  algebra  contractions in their original 2d form 
  while treating the arguments of the fields as points in $2-2 \eps$ dimensional space,
thus continuing  the  remaining scalar propagator factors to 
  AdS$_{\rd}$  with $\rd=1+d=2-2 \eps$, $\eps \to 0$.\foot{Note that  there is only  trivial  Euler number 
  1-loop divergence (that should    be cancelled by measure) 
  at the 1-loop order  \ci{Drukker:2000ep,Giombi:2020mhz}.
   Compared to  the S-matrix case   considered in \ci{Beccaria:2025xry}   here 
the   evanescent 1-loop  ${1\ov \eps} \int R^{(\rd)} $ counterterm does not contribute to  the 2-loop
free energy:   expanded  near \adst  it does not contain a non-trivial   linear in $\del x \del x$   term.}

     All 2-loop contributions will be expressed in terms of the 
regularized values of the  following   constant  factors representing  the coincident  values of the AdS$_{1+d}$ 
propagators  for a boson with mass $m^{2}_\rmb=2$ and a fermion with mass $m_\rmf=1$ \foot{Here we are not indicating  dependence on 
a normalization  mass $\mu$ and radius $r$  of AdS$_2$ 
(in general, ${1\ov \eps} \to {1\ov \eps} + 2 \log (\mu r)$) assuming  that  $ \mu r =1$.  In general, a redefinition of $\mu$   corresponds to a scheme  change (for example, we may absorb $\ell$ into $\log \mu$).} 
\ba
\la{1.8}
&\gxx =  \frac{\Gamma(\tfrac{1-d}{2})\, \Gamma(\Delta)}{(4\pi)^{(d+1)/2} \Gamma(1-d+\Delta)}=  
\frac{1}{4\pi\ept}-\frac{a_\rmb }{4\pi}+\OO ( \eps)  \ , \qquad    \Delta(\Delta-d) = m^{2}_\rmb=2 \ , \qquad   a_\rmb=2 \ , \\
&\qquad \qquad  \frac{1}{\ept} \equiv \frac{1}{\eps} + \LDR, \ \qquad   \ \ \ \LDR\equiv  \log ( 4 \pi e^{\gamma_{\rm E}}) \ , \qquad \ \  \eps= \ha ({1-d}  ) \ ,   \la{6}  \\
&\sss = - \frac{\Gamma(\frac{1-d}{2})\, \Gamma(\frac{1}{2}+\Delta )}{(4\pi)^{(d+1)/2}\Gamma(\frac{1}{2} - d + \Delta)}= 
-\frac{1}{4\pi \ept}+\frac{a_\rmf}{4\pi}+\OO ( \eps)  \ , \qquad \ \Delta= {d\ov 2} + m_\rmf , \ \ \  m_\rmf=1  \ , \quad a_\rmf=1.  \la{9}
\ea
The   massless scalars $y^a$ will 
not give a non-trivial 2-loop contribution  in dimensional regularization.\foot{Note   that we   are assuming that masses  of the fields in \rf{5} 
are kept fixed  while    replacing AdS$_{2}$  by AdS$_{2-2\eps}$. One could  consider an alternative  prescription of  keeping the  boundary dimensions of the fields  fixed to $\Delta_x=2,\,  \Delta_y=1,\,  \Delta_\theta = {3\ov2}$   but  that  approach 
does not appear to lead to a consistent picture.}
The resulting 2-loop correction $f_2$  to  the free energy in \rf{3}    will be   found to  have the   form 
\ba
\la{1.7}
& f_2=  q_\rmb  \,\gxx^{2} +\,  q_{\rmbf} \,  \gxx\, \sss + q_\rmf \,\sss^{2}\ , \\
 q_\rmb = \frac{9(1-d)}{2(1+d)}\ ,  &\qquad    q_{\rmbf} = \frac{24(1-d)}{(1+d)^{2}} \ , \qquad 
q_\rmf = \frac{2(1-d)(5-3d)}{(1+d)^{2}}\ . \la{17}
\ea
Here the  $\gxx^{2}$    contribution  comes from the quartic $x$-terms, the $\gxx\, \sss $  one  -- from the mixed $x^2$$\theta^2$ terms  and 
the $\sss^{2}$  one -- from the  quartic $\theta$ terms  in \rf{5}. 

As   each  of the three  terms in \rf{1.7} 
is multiplied  by $1-d= 2\ep$,   the
 double-pole  (log$^2$ UV divergences)   cancel independently in each of   the  three types of contributions.\foot{The  cancellation of the 2-loop  double-pole divergences in the bosonic contribution   is due  to 
  the   constant curvature of the AdS$_5$  space. This is a general  fact for a  2d 
  sigma model   on a constant   curvature space (see a discussion in \ci{Roiban:2007jf}).
  The same   cancellation   found also for the fermionic contribution  may be viewed as a 
    consequence of the residual global supersymmetry relating the bosonic contribution to the fermionic ones.} 
 However, the  single pole  $1\ov \eps$  term does not cancel:  
  \be \la{101}
  f_2 = - {11\ov 32\pi^2 \eptt} + f_{2, \rm fin}   + \OO(\ep) \  , \ \ \ \ \   \ \ \ \  \ 
  f_{2,\rm fin}=  {6 a_\rmb  + 16 a_\rmf  -19 \ov 32\pi^2}  \ , 
  \ee
  where ${1\ov \eptt } ={1\ov \eps} +2 \ell =     {1\ov \ept} +    \ell $ (cf. \rf{6}). 
  The  remaining log UV divergence  means the finite part  is a priori  scheme dependent  and  cannot be  directly 
   compared to the gauge-theory result   in \rf{4}.\foot{Note that 
  if we use dimensional regularization to  regularize also  the \adst    volume $V$  in \rf{3}  as  
  vol(AdS$_{1+d}$)=vol(AdS$_{2-2\eps}$)=$\pi^{d/2} \Gamma( - d/2)= - 2\pi [1 +  (2- \ell)\eps +...]$ 
  then the presence of a pole  in \rf{101}   will lead to an extra finite contribution to $f_2$. 
  However,  this will be  due  to  mixing  the  UV and IR regularizations; this  does not appear to be a consistent procedure in general.}
 
 One may try several options of how to  interpret the  result \rf{1.7},\rf{17}:
  
(i)   One may assume a   ``hard''  version of dimensional  reduction regularization 
  in which contractions of  all of the world-volume 
   indices  (both   ones of derivatives and  the spinor ones)  to be done first 
   in 2 dimensions and  after that  the resulting   scalar propagators 
  are to be  continued to $1+d = 2-2 \eps$  dimensions.  In this case  one is to  set  $d=1$  in the coefficients  in \rf{1.7} 
  first  so that 
   the  total 2-loop result  will be  just  zero. The issue   will then be  of  how to reconcile this    with  a  non-zero  value of $f_2$ 
  in the gauge theory  expression in \rf{4}. 
    In any case, the question of how to make this prescription consistent is a priori unclear.\foot{
    Note that compared to  standard  version of dimensional reduction 
  regularization \cite{Siegel:1979wq} were the main concern was to maintain   balance of degrees of freedom 
  to maintain supersymmetry   here we have  a separate complication of   how to continue  $\epsilon^{ab}$ terms.}

 (ii)  One  may   conjecture that the  2d UV pole in \rf{101} 
  should be    subtracted out or cancelled by an appropriate counterterm. 
   This interpretation was   used   in \ci{Beccaria:2025xry} in  the case of the
    2-loop correction  to scattering of  fluctuations on an  infinite GS  string in flat target space. 
   In the ``modified minimal subtraction''  (where one drops  the ${1\ov \eptt} $  term in  \rf{101}) 
     the finite part  in \rf{101}   will be given  by (see   \rf{1.8},\rf{9}) 
    \be 
    \la{122}
   f_{2, \rm fin}  \Big|_{a_\rmb =2, a_\rmf=1}  
   = {9\ov 32\pi^2}\ . \ee
     This is  
     3 times   larger than  the gauge theory coefficient in \rf{4}, 
    implying that  exact matching  would  require  some   additional finite  subtraction . 

  To try to modify the value  of the finite part in \rf{122} 
   one may   take into account   that    the   finite constants in the coincident values of the bosonic \rf{1.8} and fermionic \rf{9}
   propagators 
  in \adst  are, in general,  scheme-dependent  and thus special choices of them 
   may be required  for consistency  with underlying  2d supersymmetry 
  (cf. \ci{Beccaria:2019dju}). 
  One may   conjecture that  the required modification of the dimensional reduction regularization that leads to  \rf{1.8},\rf{9}
      is   the one 
   for which the  coincident values of the  bosonic and fermionic propagators  differ only  by  the sign\foot{
 As also emphasised in \cite{DHoker:1983xdu}, the regularization required to remove the coincident-point singularities of the bosonic and fermionic propagators is not unique. One may select a particular scheme by demanding consistency with the underlying symmetries of the theory. In the present case, one may expect that a preferred regularisation is fixed by demanding compatibility with the (non-linear) $\mathrm{OSp}(4^*|4)$ action on the worldsheet fluctuations. Similar relations  between the bosonic and fermionic propagators in \adst  that should   be a  consequence of maintaining \adst supersymmetry appeared  in \cite{Inami:1985di} (see also  
   \cite{Giombi:2020mhz}) 
and in \ci{Beccaria:2019dju}. }
   \be \la{13} 
  \gxx=- \sss= \frac{1}{4\pi\ept}
  -\frac{a_\rmb}{4\pi}+\OO ( \eps) \ ,  \qquad  {\rm i.e.}   \qquad a_\rmb  = a_\rmf=1  \ . 
  \ee
Assuming   this   ``supersymmetry relation''  \rf{13}
  we then conclude that  the  finite  part of \rf{101}    is given by 
\be \la{104} 
f_{2, \rm fin}\Big|_{a_\rmf  = a_\rmb =1} = {3  \ov 32\pi^2}  \ . \ee 
We thus observe that   the   formal subtraction of the  $1\ov \eptt$ 
 pole in \rf{101} 
  combined with a
scheme   choice implying \rf{13}   allows one to  
reproduce the gauge-theory value  of $f_2$ in \rf{4}.

\iffa 
Since   the finite part depends on a   regularization scheme both  prescriptions 
(direct subtraction of  the pole  and the renormalization prescription)   are effectively on an equal footing.\foot{Let us 
note that in line with a  remark below \rf{566} 
 in section \ref{4.2s} the above  conclusions about conditions on 
$ a_\rmf $ and $ a_\rmb$   in order to reproduce the  expected value of $f_2$ 
(i.e.  both   equal to 1  under simple subtraction  of the $1\ov \ept$   pole  or   both equal to 2 under  the renormalization prescription in \rf{11})  are not sensitive, as one can check, 
   to the value  of the  coefficient $q_{\rm bf}$   of the  mixed $\gxx\sss$   term  in \rf{1.7}.}
\fi

\

The structure of the rest of the paper is as follows. 
In section  2    we  will  discuss the expansion of the bosonic and fermionic parts of the \adss GS action    near \adst minimal surface 
 to quartic order in the fluctuation fields  fixing the bosonic static gauge and its $\kappa$-symmetry counterpart. 
 In section 3  we  will present the expressions for the bosonic and fermionic  kinetic operators  and their derivatives 
  required  in computing the 2-loop correction  in  dimensional regularization, i.e. 
  in AdS$_{\rd}$ with $\rd =2 -2 \eps$. The resulting  contributions  to the 2-loop  coefficient $f_2$  will be presented in section 4.
  Some concluding remarks will be made in section 5. 
  
  In Appendix \ref{apA} we will present some details  about the   computation of fermionic correlators. 
  The expression for the 2-loop  correction to  free energy   directly in 2 dimensions   (and specifically in the $\zeta$-function regularization) 
  will be discussed in Appendix \ref{apB}.
  In Appendix \ref{apC}  we will compute the 1-loop correction to the 2-point function of $x^i$  fluctuations. 
  A review of  strong coupling expansion of  $1\ov 2$ BPS WL in ABJM   theory  and some comments on 
  the corresponding string theory computation   will be given in 
  Appendix \ref{apD}.


\section{Expansion of  \adss  GS action near \adst minimal surface  }

\subsection{Bosonic part of the action  in static gauge}


Following \ci{Giombi:2017cqn} 
 we shall parametrize the \adss    metric as 
\be\la{21}
ds^{2}_{10} = \frac{(1+\frac{1}{4}\bm{x}^{2})^{2}}{(1-\frac{1}{4}\bm{x}^{2})^{2}}ds_{2}^{2}+
\frac{d\bm{x}\cdot d\bm{x}}{(1-\frac{1}{4}\bm{x}^{2})^{2}} +   \frac{d\bm{y}\cdot d\bm{y}}{(1+\frac{1}{4}\bm{y}^{2})^{2}}\ , 
\ee
where $\bm{x}=(x^{i}), \,  i= 1,2,3; \  \bm{y}=(y^{a}),\,  a= 1,...,5$  and $ds_{2}^{2}$   is the 
\adst  metric. The result of our computation will  depend on explicit  parametrization or topology 
of \adst   factor only through the   value of the volume $V$ factor so that  we may    choose it, e.g., 
  to be the Poincare half-plane (with metric  ${1\ov z^2} ( - dx_0^2 + dz^2)$) having  infinite   line  as its  boundary.  The  
minimal surface corresponding to the Wilson line at the boundary is represented by 
\ba
\la{2.2}
& x^{0}=\xi^0, \qquad z= \xi^1 \ ,  \qquad x^{i}=0, \qquad y^{a}=0, \ \ \  \ \ \   \xi^\mu = (\tau, \sigma) \ , \\
& ds_{2}^{2}=g_{\mu\nu}(\xi)\,d\xi^{\mu}d\xi^{\nu}=\frac{1}{\sigma^{2}}(-d\tau^{2}+d\sigma^{2}), \qquad g_{\mu\nu}(\xi) = \frac{1}{\sigma^{2}}\eta_{\mu\nu} \ . \la{2.3} 
\ea
The  bosonic part of the  \adss action in the static gauge  where the  fluctuations of $z$ and $x^0$ are set to  zero   has the form 
\ba
\la{2.1}
& S_{\rm b} = -T\, \int d^{2}\xi\, \sqrt{-\h} \equiv T\int d^{2}\xi\,\sqrt{-g}\, (-1 + \mc L_{\rm b}) \ , \\
& \h_{\mu\nu}=
\frac{(1+\frac{1}{4}\bm{x}^{2})^{2}}{(1-\frac{1}{4}\bm{x}^{2})^{2}}g_{\mu\nu}(\xi)
+  \frac{\partial_{\mu}\bm{x}\cdot\partial_{\nu}\bm{x}}{(1-\frac{1}{4}\bm{x}^{2})^{2}}+\frac{\partial_{\mu}\bm{y}\cdot\partial_{\nu}\bm{y}}
{(1+\frac{1}{4}\bm{y}^{2})^{2}} \ . \la{25}
\ea
The action has global  $SO(2,1)\times[SO(3)\times SO(5)]$  symmetry. Expanding in powers  of the fields 
we find the following interacting Lagrangian for the  3   massive and 5  massless 
 fluctuation fields 
\ba
\mc L_{\rm b}&=\mc L_{2}+ \mathcal{L}_{4\rm b} +\cdots, \qquad \qquad
\mathcal{L}_{4\rm b}  =  \mc L_{4x}+\mc L_{2x,2y}+\mc L_{4y}\ , \la{26} \\
\mc L_{2} &= -\tfrac{1}{2}(\partial \bm x\cdot \partial \bm x)-\bm x^{2}-\tfrac{1}{2}(\partial \bm y\cdot\partial \bm y),\la{27}  \\
\la{2.11}
\mc L_{4x} &= -\tfrac{1}{8}(\partial \bm x\cdot\partial \bm x)^{2}
+\tfrac{1}{4}(\partial x^{i}\partial x^{j})(\partial x^{i}\partial x^{j})
-\tfrac{1}{4}\bm x^{2}(\partial \bm x\cdot \partial \bm x)-\tfrac{1}{2}(\bm x^{2})^{2}, \\
\la{2.12}
\mc L_{2x,2y} &= -\tfrac{1}{4}(\partial \bm x \cdot\partial \bm x)(\partial \bm y\cdot\partial \bm y)
+\tfrac{1}{2}(\partial x^{i}\partial y^{a})(\partial x^{i}\partial y^{a}), \\
\mc L_{4y} &= \tfrac{1}{4}y^{2}(\partial \bm y\cdot\partial \bm y)-\tfrac{1}{8}(\partial \bm y\cdot\partial \bm y)^{2}
+\tfrac{1}{4}(\partial y^{a}\partial y^{b})(\partial y^{a}\partial y^{b}), \la{210}
\ea
where all   derivatives are contracted with the \adst metric.

\subsection{Fermionic part of  the  action in   static $\kappa$-symmetry  gauge \la{s2f}}

Here we  shall follow the same  notation as in \cite{Roiban:2007jf}  using Lorentzian signature  and define  the following  combinations 
of 10d  Dirac  matrices (with indices $A=(a,a')$,  $a=0,..,4;\  a'=5,..,9; i=1,2,3$) 
\begin{align}
    &\{ \Gamma_{A} , \Gamma_{B} \} = 2\eta_{AB} \ , \qquad\qquad  \eta_{AB} = (-1, +1, \dots, +1), \qquad \G_{11}= - \G_0 ...\G_9 \ , \quad \G_{11}^2 =1 \ , \la{221}  \\
    &\Gamma_{*} = i\Gamma_{0}\Gamma_{1}\Gamma_{2}\Gamma_{3}\Gamma_{4} \ , \quad \Gamma_{*}^2=1 \ ,  \quad \Gamma_{*}\Gamma_{a} = \Gamma_{a}\Gamma_{*} \
     , \quad \Gamma_{*} \Gamma_{a'} = -\Gamma_{a'}\Gamma_{*} \ ,\la{2203} \\
    &\Gamma_{*}'=i\Gamma_{5}
    \Gamma_{6}\Gamma_{7}\Gamma_{8}\Gamma_{9} \ , \qquad \Gamma_{*}'^2 = -1 \ ,\qquad  \Gamma_{11} = \Gamma_{*}\Gamma_{*}' = -\Gamma_{*}'\Gamma_{*} \ . \la{222} 
\end{align}
The two GS   fermions  $\theta^I$   ($I=1,2$)  satisfy the  Majorana-Weyl conditions
\begin{align}
    \la{2141} 
    \theta^I = \Gamma_{11}\theta^I \ , \qquad \bar{\theta}^I =(\theta^I)^{\dagger}\Gamma^{0} =  (\theta^I)^{T} \mC  \ , \qquad 
     \mC^{T} = -\mC \ , \qquad \Gamma_{A} = - \mC^{-1}\Gamma_{A}^{T}\mC \ .
\end{align}
One may choose a representation of Dirac matrices in which $\mC=\Gamma^{0}$ and  $\theta^I$ are thus  real. 
The  10d Majorana fermions $\theta, \theta' $  satisfy 
\be \label{215}
    \bar{\theta} \Gamma_{A_{1}\dots A_{n}}\theta' = (-1)^{n(n+1)\ov 2}{\bar\theta}'\Gamma_{A_{1}\dots A_{n}}\theta \ , \qquad 
    \bar{\theta} \Gamma_{B}\Gamma_{*}\Gamma_{A}\theta = -\bar{\theta}\Gamma_{A}\Gamma_{*}\Gamma_{B}\theta \ .
\ee
In the static gauge we   will split the target space  coordinates $X^M$   in 2 + 3 + 5   way (cf. \rf{2.2}) 
 here   labelling them as (interchanging 1 and 4  directions compared to \rf{21})
\be\
    X^{M} =(\xi^{0}, x^{i}, \xi^1;  y^{a'}) \ , \qquad i = 1,2, 3 \ , \qquad a' = 5, \dots, 9 \ .   \la{216}\ee
 We  also split the  corresponding tangent-space  components of 10d Dirac matrices as $\G_\hata, \G_i, \G_{a'} $ defining ``parallel''  components $\G_\hata$ as\foot{We will always assume that $\G_A$ matrices  have tangent-space indices, i.e. are constant.  Here  $\G^0 =-\G_0$ and $\epsilon^{01}= 1$. We will use the Greek letters $\a,\b, ..$ for 2d world-sheet   indices  and $\halpha, \hat \beta, ...$ for the  corresponding tangent-space indices. }
 \ba
& \Gamma_{\halpha} = (\Gamma_{0}, \Gamma_{4})   \ , \quad 
  \qquad 
\ \ \  \hGamma = \Gamma_{0}\Gamma_{4} \ , \qquad \hGamma^2 = 1 , \qquad  \Gamma^{\halpha}\Gamma^{\hbeta} = \eta^{\halpha\hbeta}-\epsilon^{\halpha\hbeta}\hGamma \ ,\la{2166} \\
&\hGamma \Gamma_{\halpha} = -\Gamma_{\halpha}\hGamma \ , \qquad \hGamma \Gamma_{i} = \Gamma_{i}\hGamma \ , \qquad \hGamma\Gamma_{a'} = \Gamma_{a'}\hGamma \ ,\la{2167} \\
&    \Gamma_{123} = \Gamma_{1}\Gamma_{2}\Gamma_{3} \  ,  \qquad 
      \la{217} 
    \Gamma_{\alpha}\Gamma_{123}= -\Gamma_{123}\Gamma_{\alpha} \ , \qquad \Gamma_{i} \Gamma_{123} = \Gamma_{123}\Gamma_{i} \ , \qquad \Gamma_{a'}\Gamma_{123} = -\Gamma_{123}\Gamma_{a'} \ , \\
    &\Gamma_{*} = -i\hGamma\Gamma_{123} = -i\Gamma_{123}\hGamma \ , \qquad \hGamma \Gamma_{*} =\Gamma_{*}\hGamma = -i\Gamma_{123} \ , 
    \qquad  \hGamma^{T}\mC = \Gamma_{4}^{T}\Gamma_{0}^{T}\mC = -\mC\hGamma \ .\la{2177} \ea
The  Minkowski signature  GS Lagrangian may be   written as  ($I,J=1,2$) 
\begin{align}
   & \L= -\sqrt{-h}-2i\epsilon^{\a\b}\int_{0}^{1}ds \, L^{A}_{\a s}\big{(}\ss^{IJ}\bar{\theta}^{I}\Gamma_{A}L^{J}_{\b s} \big{)}
     \ ,\la{2221}\\
     & 
     h_{\alpha \beta} = L_{\alpha}^{A}L_{A\beta}\ , \ \qquad \ss^{IJ}=(1, -1), \qquad \ \ L^{A}_{\alpha} = (L^{A}_{\alpha})_{s=1} \ ,
     \qquad    L^{I}_{\alpha} = (L^{I}_{\alpha})_{s=1} \ , \la{2222} 
\\
    &L^{A}_{s\a} = E^{A}_M\del_\a X^M - 4i\bar{\theta}^{I}\Gamma^{A}\big{[}\frac{\sinh^2(\tfrac{s}{2}\mathcal{M})}{\mathcal{M}^2}\big{]}^{IJ}D_\a\theta^{J} \ , \qquad \ \ \ \ \ \ L^{I}_{s\a} = \big{[} \frac{\sinh(s\mathcal{M})}{\mathcal{M}}D_\a\theta \big{]} ^{I} \ , \la{2235} \\
    & D_\a\theta^{I} = \mathcal{\nabla }_\a\theta^{I} - \frac{i}{2}\epsilon^{IJ}E^{A}_M \del_\a X^M 
    \Gamma_{*}\Gamma_{A}\theta^{J} \ , \qquad \ \ \mathcal{\nabla}_\a\theta^{I} = \del_\a\theta^{I} + \frac{1}{4}\Omega^{AB}_M \del_\a X^M \Gamma_{AB}\theta^{I} \ ,\la{2233} \\
    &(\mathcal{M}^2)^{IJ} = -\epsilon^{IK}\Gamma_{*}\Gamma^{A}\theta^{K}\bar{\theta}^{J}\Gamma_{A}+\frac{1}{2}\big(-\Gamma^{ab}\theta^{I}\bar{\theta}^{K}\Gamma_{ab}\Gamma_{*}+\Gamma^{a'b'}\theta^{I}\bar{\theta}^{K}\Gamma_{a'b'}\Gamma_{*}'\big)\epsilon^{KJ} \ .
\la{2234}  \end{align}
We will fix the $\kappa$-symmetry in  the following static-gauge  adapted  way (``$\kappa$-static'' gauge)\foot{The same gauge was used  in other contexts, e.g., in  
\ci{Kallosh:1997ky,Kallosh:1997sw,Seibold:2024oyr,Beccaria:2025xry}. } 
\be\la{255} 
    \hGamma \theta^{1} = \theta^{1} \ , \qquad\qquad  \hGamma\theta^{2} = -\theta^{2} \ , \qquad \qquad 
    \hGamma = \Gamma_{0}\Gamma_{4} \ , \quad \hGamma^2 = 1\ .
    \ee
 Thus  $\theta^1$ and $\theta^2$ will have opposite   chirality with  respect to  $\hGamma$, so that  (index $p$ stands for $i$ or $a'$) 
  \be \
  \la{256}  
  \bar{\theta}^{I}\Gamma_{p_{1}}...\Gamma_{p_{n}}\theta^{I} = 0 \ , 
\qquad    \qquad \ \ \ \bar{\theta}^{1}\Gamma_{\halpha_{1}}...\Gamma_{\halpha_{n}}\theta^{2} = (-1)^n\bar{\theta}^{1}\Gamma_{\halpha_{1}}...\Gamma_{\halpha_{n}}\theta^{2} \ .
\ee
Let us define 
\begin{align}
    &\theta \equiv \theta^{1}+\theta^2 \ , \qquad \
    \ \ \  \theta^{1} = \Pf\, \theta \ , \qquad \theta^{2} = (1-\Pf)\theta \ ,\qquad \qquad \Pf\equiv \ha ( 1 + \hGamma)  \ . \la{229}
\end{align}
\iffa 
We note that: 
\begin{align}
    &\delta^{IJ}\bar{\theta}^{I}\mathcal{A}\theta^{J} = \bar{\theta}^{1} \mathcal{A}\theta^{1}+\bar{\theta}^2\mathcal{A}\theta^2 = \bar{\theta}\mathcal{A}\theta \  , \quad \mathrm{if} \ \{\mathcal{A}, \Gamma\} = 0 \ , \\
    &s^{IJ}\bar{\theta}^{I}\mathcal{A}\theta^{J} = \bar{\theta}^{1} \mathcal{A}\theta^{1}-\bar{\theta}^2\mathcal{A}\theta^2 = \bar{\theta}\mathcal{A}\Gamma\theta \ ,  \quad \mathrm{if} \ \{\mathcal{A}, \Gamma\} = 0 \ , \\
    &\epsilon^{IJ}\bar{\theta}^{I}\mathcal{B}\theta^{J} = \bar{\theta}^{1}\mathcal{B}\theta^{2}-\bar{\theta}^{2}\mathcal{B}\theta^{1} = -\bar{\theta}\mathcal{B}\Gamma \theta \ , \quad \mathrm{if} \ [\mathcal{B}, \Gamma] = 0 \ , \\
    &s^{IL}\epsilon^{LJ}\bar{\theta}^{I}\mathcal{B}\theta^{J} = \bar{\theta}^{1}\mathcal{B}\theta^{2}+\bar{\theta}^{2}\mathcal{B}\theta^{1} = \bar{\theta}\mathcal{B}\theta \ , \quad \mathrm{if} \ [\mathcal{B}, \Gamma] = 0 \ .
\end{align}
\fi
Our aim is to expand the Lagrangian \rf{2221} to quartic order in  the independent MW   variable $\theta$. 
Since  we will be interested  only   up to quartic terms in bosons and fermions  we will use that  
in the $\theta^4$ terms we  may replace the vielbein 1-form  $E^A$   by  its \adst  value  using that  according to \rf{21}  ($e^\halpha$ is the \adst 1-form, cf. \rf{233}) 
\ba
    &E^{\halpha} = \frac{(1+\tfrac{1}{4}x^2)}{(1-\tfrac{1}{4}x^2)}e^{\halpha} \ , 
    \qquad E^{i} = \frac{dx^{i}}{(1-\tfrac{1}{4}x^2)} \ , \qquad E^{a'} = \frac{dy^{a'}}{(1+\tfrac{1}{4}y^2)} \  , \la{235}\\
     & \Omega^{\halpha \hbeta} = \omega^{\halpha \hbeta} \ , \quad 
     \Omega^{\halpha i} = \frac{e^{\halpha}x^{i}}{(1-\tfrac{1}{4}x^2)} \ , \quad \Omega^{ij} = 
      -\tfrac{1}{2}\frac{x^{i}dx^{j}-x^{j}dx^{i}}{1-\tfrac{1}{4}x^2}\ , \quad 
    \Omega^{a'b'} =  \tfrac{1}{2}\frac{y^{a'}dy^{b'}-y^{b'}dy^{a'}}{1+\tfrac{1}{4}y^2} .\la{238}
\ea
Then 
we get for  the $\hat \a$ and $p=(i, a')$   components of the 
  1-forms $L^A \equiv L^A_M  d X^M$ 
\begin{align}
    L^{\halpha}_{s} = & E^{\halpha}-is^2(\bar{\theta}^{I}\Gamma^{\halpha}D\theta^{I})-\tfrac{s^4}{12}(\bar{\theta}\Gamma_{123}\theta)(\bar{\theta}\Gamma^{\halpha}\mathfrak{D}\theta) \nonumber \\
    &-\tfrac{is^4}{24}(\bar{\theta}\Gamma^{\halpha}\Gamma^{ij}\theta)(\bar{\theta}\Gamma\Gamma_{ij}\Gamma_{*}\mathfrak{D}\theta) + \tfrac{is^4}{24}(\bar{\theta}\Gamma^{\halpha}\Gamma^{a'b'}\theta)(\bar{\theta}\Gamma\Gamma_{a'b'}\Gamma'_{*}\mathfrak{D}\theta) +  \OO(\theta^6) \ , \la{230} \\
     L^{p}_{s} =  &E^{p} - is^2(\bar{\theta}^{I}\Gamma^{p}D\theta^{I}) + \OO(\theta^6) \ ,\qquad \ \ \  E^A= E^A_M  dX^M  \la{231} \ , \\
      \mathfrak{D}\theta \equiv & \nabla \theta -\tfrac{i}{2}e^{\halpha}\hGamma\Gamma_{*}\Gamma_{\halpha}\theta = \nabla\theta +\tfrac{1}{2}e^{\halpha}\Gamma_{\halpha}\Gamma_{123}\theta \ , \la{2339} 
\end{align}
where  we used that 
$ \bar{\theta}\Gamma_{*}\theta =0, \  
     \  \epsilon^{JI}\bar{\theta}^{J}\Gamma_{\hat\a}\Gamma_{*}\Gamma_{\hat\beta}\theta^{I} = i\eta_{\hat\a\hat\b}\bar{\theta}\Gamma_{123}\theta ,\ \   \bar{\theta}^{I}\Gamma_{\halpha}\Gamma_{\hat \beta \hat\gamma}\theta^{I} =  0 .$
Expanding in powers of the bosonic fields using \rf{235},\rf{238}  we have 
\begin{align}
    \bar{\theta}^{I}\Gamma^{ \halpha}D\theta^{I} &=\te  \bar{\theta}^{I}\Gamma^{ \halpha}(\nabla\theta^{I}-\frac{i}{2}\epsilon^{IJ}E^{\hbeta}\Gamma_{*}\Gamma_{\hbeta}\theta^{J}
    +\frac{1}{4}\Omega^{ij}\Gamma_{ij}\theta^{I}+\frac{1}{4}\Omega^{a'b'}\Gamma_{a'b'}\theta^{J}) \nonumber \\
    &\te  = \bar{\theta}\Gamma^{ \halpha}\mathfrak{D}\theta -\frac{1}{4}x^2 e^{\hbeta}(\bar{\theta}\Gamma^{ \halpha}\Gamma_{123}\Gamma_{\hbeta}\theta)  - \frac{1}{4}x^{i}dx^{j}(\bar{\theta}\Gamma^{ \halpha}\Gamma_{ij}\theta) +\frac{1}{4}y^{a'}dy^{b'}(\bar{\theta}\Gamma^{ \halpha}\Gamma_{a'b'}\theta)+\dots \ , \la{2348}\\
    \te \bar{\theta}^{I}\Gamma^{p}D\theta^{I} &\te= \bar{\theta}^{I}\Gamma^{p}(\frac{1}{2}\Omega^{ \halpha j}\Gamma_{ \halpha}\Gamma_{j}\theta^{I}-\frac{i}{2}\epsilon^{IJ}E^{j}\Gamma_{*}\Gamma_{j}\theta^{J}-\frac{i}{2}\epsilon^{IJ}E^{a'}\Gamma_{*}\Gamma_{a'}\theta^{I}) \nonumber \\
    & \te = \frac{1}{2} e^{ \halpha}x^{j}(\bar{\theta}\Gamma^{p}\Gamma_{ \halpha}\Gamma_{j}\theta)+\frac{i}{2}dx^{j}(\bar{\theta}\Gamma\Gamma^{p}\Gamma_{*}\Gamma_{j}\theta) + \tfrac{i}{2}dy^{a'}(\bar{
    \theta
    }\Gamma\Gamma^{p} \Gamma_{*}\Gamma_{a'}\theta)+\dots \ . \la{2347}
\end{align}
Using  the  \adst 2-bein  $e^{\halpha}_\mu$   we may define the world-volume   projectors of $\G_\a$ as 
\be 
\g_\mu \equiv e^\hta_\mu  \G_\hta \ , \ \ \ \ \ \ \ \   \{\g_\hta, \g_\htb\} = 2 g_{\hta\htb} \ , \ \ \ \ \ 
g_{\m\n} = e^\hta_\m e^\htb_\nu \eta_{\hta\htb} \ . \la{233} \ee
Then  we find  that 
\begin{align}
    h_{\alpha\beta}  =&\te  L^{A}_{\alpha}L_{A\beta}  = E^{A}_{\alpha}E^{A}_{\beta}-2i(1+\frac{1}{2}x^2) (\bar{\theta}\gamma_{(\alpha}\mathfrak{D}_{\beta)}\theta) +\frac{i}{2}x^2\bar{\theta}\gamma_{(\alpha}\Gamma_{123}\gamma_{\beta)}\theta  \nonumber \\
    &\te+\frac{i}{2}x^{i}\partial_{(\alpha}x^{j}(\bar{\theta}\gamma_{\beta)}\Gamma_{ij}\theta)-\frac{i}{2}y^{a'}\partial_{(\alpha}y^{b'}(\bar{\theta}\gamma_{\beta)}\Gamma_{a'b'}\theta)- (\bar{\theta}\gamma^{\mu}\mathfrak{D}_{\alpha}\theta)(\bar{\theta}\gamma_{\mu}\mathfrak{D}_{\beta}\theta) \nonumber \\
    & -iE^{p}_{(\alpha}\Big{[} x^{j}(\bar{\theta}\Gamma^{p}\gamma_{\beta)}\Gamma_{j}\theta) + \partial_{\beta)}x^{j}(\bar{\theta}\Gamma^{p}\Gamma_{123}\Gamma_{j}\theta)   + \partial_{\beta)}y^{a'}(\bar{\theta}\Gamma^{p}\Gamma_{123}\Gamma_{a'}\theta)\Big{]} \la{2370} \\
    &\te -\frac{1}{6}(\bar{\theta}\Gamma_{123}\theta)(\bar{\theta}\gamma_{(\alpha}\mathfrak{D}_{\beta)}\theta) - \frac{i}{12}(\bar{\theta}\gamma_{(\alpha}\Gamma^{ij}\theta)(\bar{\theta}\hGamma\Gamma_{ij}\Gamma_{*}\mathfrak{D}_{\beta)}\theta)+\frac{i}{12}(\bar{\theta}\gamma_{(\alpha}\Gamma^{a'b'}\theta)(\bar{\theta}\hGamma\Gamma_{a'b'}\Gamma'_{*}\mathfrak{D}_{\beta)}\theta) + \dots . \no
\end{align}
Thus  the  first term in \rf{2221} is given by 
\begin{align}
    & -\sqrt{-h} =\te  -\sqrt{-\h} +i\sqrt{-\h}\, \h^{\alpha\beta}\bar{\theta}\gamma_{\alpha}\mathfrak{D}_{\beta}\theta + \sqrt{-g}\Big{[}\frac{i}{2}x^2 \bar{\theta}\slashed{\mathfrak{D}}\theta +\frac{i}{2}x^2\bar{\theta}\Gamma_{123}\theta  
    -\frac{i}{4}(x^{i}\partial_{\alpha}x^{j})(\bar{\theta}\gamma^{\alpha}\Gamma_{ij}\theta)\no
    \\
    &\te +\frac{i}{4}(y^{a'}\partial_{\alpha}y^{b'})(\bar{\theta}\gamma^{\alpha}\Gamma_{a'b'}\theta) 
    +\frac{1}{2}(\bar{\theta}\gamma^{\rho}\mathfrak{D}_{\alpha}\theta)(\bar{\theta}\gamma_{\rho}\mathfrak{D}^{\alpha}\theta) + \frac{1}{2}(\bar{\theta}\slashed{\mathfrak{D}}\theta)^2-g^{\alpha\beta}g^{\gamma\delta}(\bar{\theta}\gamma_{(\alpha}\mathfrak{D}_{\gamma)}\theta)(\bar{\theta}\gamma_{(\beta}\mathfrak{D}_{\delta)}\theta) \nonumber \\
    &\te +\frac{1}{12}(\bar{\theta}\Gamma_{123}\theta)(\bar{\theta}\slashed{\mathfrak{D}}\theta) + \frac{i}{24}(\bar{\theta}\gamma^{\alpha}\Gamma^{ij}\theta)(\bar{\theta}\Gamma \Gamma_{ij}\Gamma_{*}\mathfrak{D}_{\alpha}\theta)-\frac{i}{24}(\bar{\theta}\gamma^{\alpha}\Gamma^{a'b'}\theta)(\bar{\theta}\Gamma \Gamma_{a'b'}\Gamma'_{*}\mathfrak{D}_{\alpha}\theta) \nonumber \\
    &\te  
    +\frac{i}{2}\partial^{\alpha}x^{i}\partial_{\a}x^{j}\, \bar{\theta}\Gamma_{i}\Gamma_{123}\Gamma_{j}\theta \Big{]}+ \dots \ . \la{2390}
\end{align}
Here $\h_{\a\b}$ is the  full bosonic   part of the induced metric in \rf{25} 
 while $g_{\a\b}$ is the \adst metric, so that (cf. \rf{26})
\begin{align}
 \te    \sqrt{-\h}\, \h^{\alpha\beta} =  \sqrt{-g}\Big[g^{\alpha\beta}-\partial^{\alpha}x^{i}\partial^{\beta}x^{i}-(\partial^{\alpha}y^{a'})(\partial^{\beta}y^{a'})+\frac{1}{2}g^{\alpha\beta}\big((\partial x)^2+(\partial y)^2\big)+\dots\Big] \ .\la{4000}
\end{align}
To find the  contribution of the WZ  term in \rf{221}  we  note that (see \rf{2235})  
\begin{align}
    \ss^{IJ}\bar{\theta}^{I}\Gamma^{\halpha}L_{s}^{J}=&\te s\, \ss^{IJ}\bar{\theta}^{I}\Gamma^{\halpha}D\theta^{J}-\frac{is^3}{6}\epsilon^{\halpha\hbeta}(\bar{\theta}\Gamma_{123}\theta)(\bar{\theta}\Gamma_{\hbeta}\mathfrak{D}\theta) \nonumber \\
    &\te  +\frac{s^3}{12}(\bar{\theta}\Gamma^{\halpha}\Gamma^{ij}\hGamma\theta)(\bar{\theta}\hGamma\Gamma_{ij}\Gamma_{*}\mathfrak{D}\theta)-\frac{s^3}{12}(\bar{\theta}\Gamma^{\halpha}\Gamma^{a'b'}\hGamma\theta)(\bar{\theta}\hGamma\Gamma_{a'b'}\Gamma'_{*}\mathfrak{D}\theta)+\OO(\theta^6)  \ , \la{247}\\
     \ss^{IJ}\bar{\theta}^{I}\Gamma^{p}L_{s}^{J}=&s\, \ss^{IJ}\bar{\theta}^{I}\Gamma^{p}D\theta^{J}+\OO(\theta^6)  \ .\la{2448}
\end{align}
Thus  
\begin{align}
    &-2i \ss^{IJ} \int_{0}^{1}ds \, \epsilon^{\alpha\beta}L^{A}_{\a s} \bar{\theta}^{I}\Gamma^{A}L_{\b s}^{J} \te =\sqrt{-g} \Big\{
    i(1+\frac{1}{2}x^2+\dots)\bar{\theta}\gamma^{\beta}\Big[\mathfrak{D}_{\beta}\theta -\frac{1}{4}
    x^2\Gamma_{123}\gamma_{\beta}\theta
    \no\\  &\te\la{3303}
     \qquad \qquad   -\frac{1}{4}x^{i}\partial_{\beta} x^{j}\, \Gamma_{ij}\theta  +\frac{1}{4}y^{a'}\partial_{\beta}y^{b'}\, \Gamma_{a'b'}\theta+\dots\Big]
     +\frac{1}{12} (\bar{\theta}\Gamma_{123}\theta)(\bar{\theta}\slashed{\mathfrak{D}}\theta) \\ &\qquad\qquad   \te +\frac{i}{24} (\bar{\theta}\gamma^{\beta}\Gamma^{ij}\theta)(\bar{\theta}\hGamma\Gamma_{ij}\Gamma_{*}\mathfrak{D}_{\beta}\theta) - \frac{i}{24} (\bar{\theta}\gamma^{\beta}\Gamma^{a'b'}\theta)(\bar{\theta}\hGamma\Gamma_{a'b'}\Gamma'_{*}\mathfrak{D}_{\beta}\theta) \nonumber \\
    &\qquad \te 
    +\frac{1}{2} \Big{[} (\bar{\theta}\slashed{\mathfrak{D}}\theta)^2 - (\bar{\theta}\gamma_{\beta}\mathfrak{D}_{\alpha}\theta)(\bar{\theta}\gamma^\a\mathfrak{D}^\beta \theta)\Big{]} 
    +\frac{i}{2} x^{j}\partial_{\alpha}x^{p} ( \bar{\theta}\gamma^{\alpha} \Gamma_{p} \Gamma_{j}\theta) \Big\} - 
    \frac{1}{2}\epsilon^{\alpha\beta}\partial_{\alpha}x^{i}\partial_{\beta}x^{j}(\bar{\theta}\Gamma_{i}\Gamma_{*}\Gamma_{j}\theta)+\dots, \no 
\end{align}
where we  used that 
$ \epsilon^{\alpha\beta}(\bar{\theta}\gamma_{\delta }\mathfrak{D}_{\alpha}\theta)(\bar{\theta}\gamma^{\delta}\hGamma \mathfrak{D}_{\beta}\theta) = -\sqrt{-g}\big{[}(\bar{\theta}\slashed{\mathfrak{D}}\theta)^2 - (\bar{\theta}\gamma_{\b}\mathfrak{D}_{\alpha}\theta)(\bar{\theta}\gamma^\a\mathfrak{D}^\b\theta)\big{]} $.

As a result,  the fermionic part of the GS Lagrangian expanded to quartic order in the  fluctuation fields 
 is found 
(after rescaling $\theta \rightarrow \frac{1}{2}\theta$ and factoring out $\sqrt{-g}$) to be:
\begin{align}
\mathcal{L}_{\rm f}=  & \ha  i\bar{\theta}\slashed{\mathfrak{D}}\theta + \mathcal{L}_{\rm 4f} + ... \ , \la{404}\\
\mathcal{L}_{\rm 4f} = &
   \te \frac{1}{4}i \Big{[} x^2g^{\alpha\beta}-\partial^{\alpha}x^{i}\partial^{\beta}x^{i}-\partial^{\alpha}y^{a'}\partial^{\beta}y^{a'}+\frac{1}{2}g^{\alpha\beta}\big((\partial x)^2+(\partial y)^2\big)\Big{]} \bar{\theta}\gamma_{\alpha}\mathfrak{D}_{\beta}\theta \nonumber \\
    &\te +\frac{i}{8}\big{(} \partial^{\alpha}x^{i}\partial_{\alpha}x^{i}+2x^2-\partial^{\alpha}y^{a'} \partial_{\alpha}y^{a'}\big{)}\bar{\theta}\Gamma_{123}\theta \nonumber \\
    &\te +\frac{1}{96}(\bar{\theta}\Gamma_{123}\theta)(\bar{\theta}\slashed{\mathfrak{D}}\theta)+\frac{i}{192}(\bar{\theta}\gamma^{\a}\Gamma^{ij}\theta)(\bar{\theta}\hGamma\Gamma_{ij}\Gamma_{*}\mathfrak{D}_{\a}\theta) - \frac{i}{192}(\bar{\theta}\gamma^{\a}\Gamma^{a'b'}\theta)(\bar{\theta}\hGamma\Gamma_{a'b'}\Gamma'_{*}\mathfrak{D}_{\a}\theta) \nonumber \\
    &\te +\frac{1}{16} (\bar{\theta}\slashed{\mathfrak{D}}\theta)^2 - \frac{1}{16} 
     (\bar{\theta}\gamma^{\alpha}\mathfrak{D}^{\b}\theta)(\bar{\theta}\gamma_{\b}\mathfrak{D}_{\alpha}\theta)  +\dots \ ,  \la{2431}\\
     &\te  \mathfrak{D}_\a\theta  
     \equiv  \nabla_\a \theta +\frac{1}{2}\gamma_{\alpha}\Gamma_{123}\theta \ ,\qquad \ \ \ \ 
    \slashed{\mathfrak{D}} = \slashed{\nabla}+\Gamma_{123} \ .\la{2455}
\end{align}
Note that we have dropped  the   terms  in \rf{2390},\rf{3303} 
\be 
\Delta \mc L=  -\tfrac{i}{2}(x^{i}\partial_{\alpha}x^{j})(\bar{\theta}\gamma^{\alpha}\Gamma_{ij}\theta)  +\tfrac{i}{2}(y^{a'}\partial_{\alpha}y^{b'})(\bar{\theta}\gamma^{\alpha}\Gamma_{a'b'}\theta) 
-
    \tfrac{1}{2\sqrt {-g} }\epsilon^{\alpha\beta}\partial_{\alpha}x^{i}\partial_{\beta}x^{j}(\bar{\theta}\Gamma_{i}\Gamma_{*}\Gamma_{j}\theta)\ , \la{5566}
\ee
 which  will  not contribute to the  2-loop 
free energy.  


\section{AdS propagators}

 Below  we   will use the dimensional reduction regularization  by replacing \adst with  AdS$_{\rd}$ with $\rd=1+d, \, d=1-2 \ep$
 but  not changing the number of fermionic  components  and  keeping $\rd=2$ in gamma  matrix contractions.
  Let us first review  the standard expressions for
   the  scalar and fermion propagators  in AdS$_{1+d}$.

\subsection{Scalar propagator in AdS$_{1+d}$}

In Euclidean AdS$_{1+d}$ with the Poincare  coordinates $\w^{\mu}=(z, w^{r})$  
 we have
for a massive scalar field with the standard Dirichlet boundary conditions (see, e.g.,  \cite{DHoker:2002nbb})\foot{The 
Dirichlet conditions  correspond to a  supersymmetric (Maldacena-Wilson loop)   case. If one uses the  Neumann    boundary conditions 
for the $S^5$ massless modes  that would correspond to the case of the standard (non-BPS) WL \ci{Beccaria:2019dws}.} 
\ba\la{32}
&\hat S = \frac{1}{2}\int d^{d+1}\w\, \sqrt{g}(g^{\mu\nu} \del_{\mu}\phi\del_{\nu}\phi+m^{2}\phi^{2}), \qquad
ds^2 = g_{\mu \nu }(\w) d \w^\mu d \w^\nu =  {1\ov z^2} ( dz^2 + dw_r dw^r)\ , \\
&\langle \phi(\w)\phi(\w') \rangle = G_{\Delta}(\w,\w'), \qquad \Delta = \frac{d}{2}+ \sqrt{\frac{d^{2}}{4}+m^{2}},\ \ \ \ \ 
\Delta (\Delta - d)= m^2 \ , 
\la{33} \\
& (-  \nabla^2 + m^2)    G_{\Delta}(\w, \w')  =   \frac{1}{\sqrt g}\delta^{(d+1)}(\w-\w') \ , \la{337}
\\
\la{4.6}
&G_{\Delta}(\w, \w') = \frac{C_{\Delta}}{2\Delta-d}\Big[\frac{v(\w,\w')}{2}\Big]^{\Delta}{}_{2}F_{1}\Big(\frac{\Delta}{2}, \frac{\Delta}{2}+\frac{1}{2};\Delta-\frac{d}{2}+1; v^{2}(\w,\w')\Big),
\\
&C_{\Delta} = \frac{\Gamma(\Delta)}{\pi^{d/2}\Gamma(\Delta-\frac{d}{2})} \ , \qquad\qquad 
 v(\w,\w') \equiv \frac{2zz'}{z^{2}+z'^{2}+(w^{r}-w'^{r})^{2}}  \ . \la{35}
\ea
In Minkowski  AdS$_{1+d}$  
we get the same propagator   with the replacement  of the   $\delta_{rs}$ metric in $v$ in \rf{35}  by the Minkowski one. 
Using 
that 
\be
\la{4.11}
_{2}F_{1}(a,b,c; 1) = \frac{\Gamma(c)\Gamma(c-b-a)}{\Gamma(c-b)\Gamma(c-a)}, \qquad \Re(c-a-b)>0, \ \ c\neq 0, -1, -2, \dots,
\ee
and  doing the analytic continuation in $d$ to extend beyond the above conditions on the  parameters $a,b,c$   we get 
for the scalar  propagator at  the coinciding points 
\be
\la{4.12}
G_{\Delta}(\w,\w) = \frac{1}{(4\pi)^{(d+1)/2}}\frac{\Gamma(\tfrac{1}{2}-\tfrac{d}{2})\, \Gamma(\Delta)}{\Gamma(1-d+\Delta)}.
\ee
The first derivatives of the propagator  taken at the coinciding points vanish, 
$\partial_{\mu}G_{\Delta}(\w,\w') \big|_{\w=\w'} =0$  (note that  for $v$ in \rf{35} one has $\partial_{\mu} v(\w,\w') \big|_{\w=\w'} =0$). 
We  need also the  value of the  second  derivative\foot{Here   and below $\partial'_{\mu} $ stands for the derivative over 
$\w'$ with  index $\mu$.} 
in  the limit $\w'\to \w$.  We get  using again (\ref{4.11})
\be
\la{4.19}
\nabla_{\mu}\nabla'^{\mu}G_{\Delta}(\w,\w') \big|_{\w=\w'} =\frac{d+1}{2}
 \frac{1}{(4\pi)^{(d+1)/2}}\frac{\Gamma(-\tfrac{1}{2}-\tfrac{d}{2})\, \Gamma(\Delta+1)}{\Gamma(-d+\Delta)} 
 = - m^2 G_{\Delta}(\w,\w)  \ , 
\ee
where we used \rf{4.12} and that $\Delta (\Delta - d)= m^2$. This relation  is in agreement with \rf{337}  and $\delta^{(1+d)}(\w,\w)=0$ in dimensional regularization. 
Similarly, one finds that 
\ba\la{310}
\nabla_{\mu}\nabla'_{\nu}G_{\Delta}(\w,\w')\big|_{\w=\w'} =
\frac{1}{2}g_{\mu\nu} \frac{1}{(4\pi)^{(d+1)/2}}\frac{\Gamma(-\tfrac{1}{2}-\tfrac{d}{2})\, \Gamma(\Delta+1)}{\Gamma(-d+\Delta)}
= - {m^2   \ov d+1}  g_{\mu\nu}  G_{\Delta}(\w,\w) \ . 
\ea
Setting $d=1-2\eps$ and expanding in small $\eps$  with  $\Delta$  defined as in \rf{33}  
we find 
\ba
\la{38} 
 G_{\Delta}(\w,\w) = \frac{1}{4\pi\eps}+\frac{1}{4\pi}\big[ \log(4\pi)-\gamma_{\rm E}-2\psi(\Delta_{0})\big]+\cdots, \qquad 
\Delta_{0}\equiv \Delta\big|_{\ep=0} = \frac{1}{2}+\sqrt{\frac{1}{4}+m^{2}}\ .  
\ea
For the special   values of  the masses  we are interested in we get 
\ba\la{312}
m^{2}=0:\ \  \Delta_{0} =1, \quad  \psi(\Delta_{0}) = -\gamma_{\rm E}\ ; \qquad \ \ 
m^{2}=2:\ \ \Delta_{0} = 2, \quad  \ \psi(\Delta_{0}) = 1-\gamma_{\rm E}.
\ea
Note that for the massless   field the expressions in \rf{4.19},\rf{310} vanish; as a result, the  correlators of the $y^a$ field in \rf{26} will not  contribute to  the 2-loop free energy.

For the  propagator of the $m^2 =2$ scalar   we will  introduce the special notation 
\be \la{333}
 \ \ \ \ \gxx (\w, \w') \equiv G_\Delta (\w,\w')\big|_{m^2=2} \ , \qquad \qquad \Delta = {d\ov 2} + \sqrt{ {d^2\ov 4}  + 2 } \ . \ee 
Expanding in $\eps= \ha (1-d) \to 0$ we get\foot{\la{400}For the massless field   with $\Delta =d$ we have from \rf{38},\rf{312}
$  G_\Delta (\w, \w)\big|_{m^2=0}  =  \frac{1}{4\pi\ept}   +\OO(\eps)$.}
\ba
\la{4.21}
\gxx\equiv  \gxx (\w, \w) &=  \frac{1}{4\pi\ept}   
-\frac{1}{2\pi}+\OO(\eps) \ , \qquad \qquad  \frac{1}{\ept} \equiv \frac{1}{\eps} + \LDR\ , 
\qquad   \LDR = \log(4\pi e^{\gamma_{\rm E}}) \ ,  \\
\nabla_{\mu}\nabla'^{\mu} \gxx(\w,\w')\big|_{\w=\w'} &= -2\gxx = -\frac{1}{2\pi\ept }  
+\frac{1}{\pi} + \OO(\eps) \la{334}  \ , 
\\
\nabla_{\mu}\nabla'_{\nu} \gxx(\w,\w')\big|_{\w=\w'} &= - {2\ov d+1} g_{\mu\nu} 
 \gxx = g_{\mu\nu}    \big[-\frac{1}{4\pi\ept}   
 +\frac{1}{4\pi} + \OO (\eps) 
  \big] \ . \la{335}
\ea

\subsection{Spinor propagator in  AdS$_{1+d}$}
Let us consider the   spinor field in Euclidean AdS$_{1+d}$  in Poincare coordinates   with the action 
\be
S_{F} = \int d^{d+1}\w\ \sqrt g\ \bar\psi(\slashed{\nabla}-M)\psi\ , \la{3166}
\ee
where $ \slashed{\nabla}=e^{\alpha}_{\hta}\gamma^{\hta}(\partial_{\alpha}+\tfrac{1}{2}\omega^{\hta\htb}_{\alpha}\Omega_{\hta\htb}), \
 \gamma^{(\hta}\gamma^{\htb)}=\delta^{\hta\htb}, \  \Omega^{\hta\htb }=\tfrac{1}{4}[\gamma^{\hta},\gamma^{\htb}]
$, $e^{\hta}_{\alpha} = \frac{1}{z}\delta^{\hta}_{\alpha},$  $\omega^{\hta\htb}_{\hat \g} = \frac{1}{z}(\delta^{\hta}_{0}\delta^{\htb}_{\hat \gamma }-\delta^{\htb}_{0}\delta^{\hta}_{\hat \gamma}). $   Explicitly (cf., e.g.,  \cite{Kawano:1999au})
\be
\la{4.32}
\slashed{\nabla} = 
z\gamma^{\hta}\partial_{\hta}-\tfrac{d}{2}\gamma_{\hat 0} \ , \qquad \qquad \del_\hta = \delta^\alpha_\hta \del_\alpha \ , 
\ee
where $\hat 0$ denotes the tangent index in the  $z$-direction.  
Let us consider the  corresponding propagator defined by 
\ba
\la{4.33}
(\slashed{\nabla}-M)S(\w,\w') = \frac{1}{\sqrt g}\delta^{(d+1)}(\w,\w') \ . 
\ea
 In the standard  case when  $M$ commutes   with $\gamma_\hta$, e.g. if  $M=m I $, 
 the propagator is given by (see, e.g., \cite{Basu:2006ti})
 \ba
&S(\w,\w') = -\frac{1}{2^{m+(d+3)/2}\pi^{d/2}}\frac{\Gamma(m+\frac{d+1}{2})}{\Gamma(m+\frac{1}{2})}\frac{1}{\sqrt{zz'}}\frac{1}{(u+2)^{m+(d+1)/2}}\no \\
&\ \qquad \qquad  \qquad  \times     \Big[(\slashed{\w}\gamma_{\hat 0}+\gamma_{\hat 0}\slashed{\w'})\, {\rm F}_{1}(u)-(\slashed{\w}-\slashed{\w'})\, {\rm F}_{2}(u)    \Big],\la{322} \\
& u(\w,\w') \equiv  \frac{(z-z')^{2}+ (w^{r}-w'^{r})^{2}}{2zz'} \ , \qquad \slashed{\w} = \gamma_{\alpha}\w^{\alpha}, \la{223} \\ 
&{\rm F}_{1}(u) = {}_{2}F_{1}(m+\tfrac{d+1}{2}, m, 2m+1, \tfrac{2}{u+2}), \qquad\ \ \ 
{\rm F}_{2}(u) = {}_{2}F_{1}(m+\tfrac{d+1}{2}, m+1, 2m+1, \tfrac{2}{u+2}).\la{307}
\ea
One can check that  if  the mass term $M$  anticommutes with $\gamma_\a$, i.e. 
\be
\la{4.38} M = - m \bar\gamma, \qquad  \qquad  \ \{\gamma^{\hta},\bar\gamma\}=0, 
\qquad \bar\gamma^{2}=-1, \qquad m > 0 \ , 
\ee
eq. \rf{4.33} is solved by  a similar matrix   function 
\ba
\la{4.39}
S(\w,\w') &= -\frac{1}{2^{m+(d+3)/2}\pi^{d/2}}\frac{\Gamma(m+\frac{d+1}{2})}{\Gamma(m+\frac{1}{2})}\frac{1}{\sqrt{zz'}}\frac{1}{(u+2)^{m+(d+1)/2}}\lp\qquad 
\times     \Big[\bar\gamma\, (\slashed{\w}\gamma_{\hat 0}+\gamma_{\hat 0}\slashed{\w'})\, {\rm F}_{1}(u)-(\slashed{\w}-\slashed{\w'})\, {\rm F}_{2}(u)    \Big] \ . 
\ea
The   same  expression is found in the case of Minkowski signature AdS$_{1+d}$   space provided one uses the Minkowski metric 
in contracting indices of $w^r$  in $u(\w,\w')$ in \rf{223}. 

In  the  GS   string   case   we are interested in (see \rf{2455})  we have \be \bar \g = \G_{123}\ , \qquad \qquad 
\gamma_{\hat 0} = \G_4 \ . \la{099}\ee
 Note that in  the special case of   $d=1$ and $m=1$  one finds explicitly 
\ba
S(\w,\w') &= -\frac{1}{4\pi}\frac{1}{\sqrt{zz'}}\frac{1}{(u+2)^{2}}
     \Big[ \Gamma_{123}(\slashed{\w}\Gamma_{4}+\Gamma_{4}\slashed{\w'})\, {\rm F}_{1}(u)
-(\slashed{\w}-\slashed{\w'})\, {\rm F}_{2}(u)    \Big],\la{329}
\\ {\rm F}_{1}(u)  &= -\tfrac{1}{2}(u+2)(2+(u+2)\log\tfrac{u}{u+2})\ , \qquad \ \ \ 
{\rm F}_{2}(u) = \tfrac{(u+2)^{2}}{2u}(2+u\log\tfrac{u}{u+2}). \la{339}
\ea
In the limit  of coinciding points, i.e. $\w=\w'$ and 
$u=0$, using (\ref{4.11}) we get from \rf{307}
\ba
{\rm F}_{1}(0) &= {}_{2}F_{1}(m+\tfrac{d+1}{2}, m, 2m+1, 1) = \frac{2^{2m}\Gamma(m+\frac{1}{2})}{\sqrt\pi}\frac{\Gamma(\frac{1-d}{2})}{\Gamma(\frac{1-d}{2}+m)}\ ,\la{133}  \\
{\rm F}_{2}(0) &= {}_{2}F_{1}(m+\tfrac{d+1}{2}, m+1, 2m+1, 1) = \frac{\Gamma(1+2m)}{\Gamma(m)}\frac{\Gamma(-\frac{d+1}{2})}{\Gamma(\frac{1-d}{2}+m)}\ . \la{2373}
\ea
 The factor in the second line in (\ref{4.39}) reduces to 
$
\Gamma_{123}\{\slashed{\w},\Gamma_{4}\}\ {\rm F}_{1}(0) \to  2 \sigma\Gamma_{123}\,\ {\rm F}_{1}(0) $
(where $\s= w^4$) 
so that we get (cf. \rf{9}) 
\be\la{3344}
 S(\w,\w) = \sss\,  \Gamma_{123}, \ \ \ \ \ \qquad 
 \sss= - \frac{1}{(4\pi)^{(d+1)/2}}\ \frac{\Gamma(\frac{1-d}{2})\Gamma(\frac{1+d}{2}+m)}{\Gamma(\frac{1-d}{2}+m)}\ .
\ee
In the case of   $m=1$   and $d=1-2\eps$   taking  $\eps\to 0 $   (and using the same  notation as  in \rf{4.21})
we find 
\be
\sss=   -   \frac{1}{4\pi\ept}+ \frac{1}{4\pi}   +\mc O(\eps). \la{99} 
\ee
We will also need  the expressions for the derivative of the spinor propagator 
\be
\nabla_{\alpha}S(\w,\w')\big|_{\w=\w'} = \frac{1}{2 (4\pi)^{(d+1)/2}}\frac{\Gamma(-\frac{d+1}{2})\Gamma(\frac{d+1}{2}+m)}{\Gamma(\frac{1-d}{2}+m)}\,  \frac{m}{z}\Gamma_{\hat \alpha} =\frac{1}{1+d}\frac{m}{z}\Gamma_{\hat \alpha}\, \sss\ .  \la{336} 
\ee
As a check,  we  may  compute $\slashed{\nabla}S(\w,\w')\big|_{\w=\w'}$  and  find that the resulting expression is consistent 
with the definition of the Dirac propagator  ($\Gamma_{123}^{2}=-1$)
\be
\slashed{\nabla}S(\w,\w')\big|_{\w=\w'} = -m\Gamma_{123}S(\w,\w) = m\,  \sss\ . \la{3371}
\ee
Note that $\sss$ in \rf{3344},\rf{99} representing the $m^2=1$ fermion propagator  at the coinciding points 
   is related to the $m^2=2$ and $m^2=0$  values of the   bosonic propagators given  in \rf{4.21} and  the footnote \ref{400}  as 
\be 
\sss = - \ha \Big[ G_\Delta (\w,\w)\big|_{m^2=2}  + G_\Delta(\w,\w)\big|_{m^2=0}  \Big] \ . \la{401} \ee
\iffa 
\begin{remark}
We have the supersymmetry relation
\be
\la{4.60}
\sss = -[\gxx'_{\frac{d}{2}+m+\frac{1}{2}}(u)-\gxx'_{\frac{d}{2}+m-\frac{1}{2}}(u)]\big|_{u=0,m=1}.
\ee
When $m=\frac{3-d}{2}$ this gives the conformal dimensions $\Delta= 2$ and $\Delta=1$. See also 
remark  in footnote 14 and (A.5) in \cite{Giombi:2020mhz}.
\end{remark}
\fi 
As was already mentioned  above, for the  correlators of the  GS fermions  we will  use  the 
  dimensional reduction   regularization prescription:
we will  not continue  the  world-volume components of   the 
gamma-matrices   to $d=1-2 \ep$   doing all  spinor  index contractions  and traces  in $1+d=2$. 
In particular, we   will use that $\gamma_\alpha \gamma^\alpha =2$ rather than $1+d$    so that \rf{3371}   will be modified to 
\be
\slashed{\nabla}S(\w,\w')\big|_{\w=\w'} = {2\ov 1+d}  m\,  \sss\ . \la{33710}
\ee

\section{2-loop contribution to string  free energy}

Our aim  is to compute  the  2-loop  coefficient $f_2$ in the string free  energy  in \rf{3},\rf{4}
starting   with the quartic terms \rf{26},\rf{2431}  in the GS  string Lagrangian. 
In the dimensional regularization  there will be no contributions from possible local measure and 
 ghost determinants  for the (algebraic)   static  bosonic and fermionic $\kappa$-symmetry gauges.

 
 \subsection{Bosonic contribution}
 
 In the Euclidean notation where   instead of \rf{2.1}  we have the  path integral   with $e^{-\hat S} $  where $\hat S= \hat S_{\rm b} + \hat S_{\rm f}$ and 
  $\hat S_{\rm b}=T\int d^{2}\xi\,\sqrt{g}\, (1 - \mc L_{\rm b }) = T V - T\int d^{2}\xi\,\sqrt{g}\,  \mc L_{\rm b} $. Here 
  $\L_{\rm b} = - \ha ( \del x_i)^2 + ...+ \L_{4x} + ...$  with the same signs 
  as in \rf{26}    but now with the  Euclidean $g_{\m\n}$ 
  comparing to \rf{4}. Then    we  have  for the 2-loop coefficient in \rf{3},\rf{4} 
  \be \la{319} 
  f_2 =  \langle \L_{4x} + ...   \rangle  \ . \ee 
  Here   dots stand for other quartic  vertices. 
  We used   that the expectation  value  of the Lagrangian 
   (computed with the  Euclidean \adst propagator $ \langle  x x   \rangle $ after rescaling by   $T$)
      will be constant so the  \adst volume $V$  can be   factored  out. 

Since  the  $y^a$ dependent terms in the  Lagrangian in \rf{26}--\rf{210}  always   contain   factors 
with  derivatives  and since in dimensional regularization  the derivatives of the massless field  propagator  taken at coinciding points vanish (cf. \rf{310})   only the  $m^2=2$ field $x^i$  correlators will contribute  to $f_2$ in \rf{319}. 

Let us  add for   generality some coefficients $b_k$ in front of  each quartic  term  in  (\ref{2.11}), i.e.  consider 
\ba\la{041} \mc L_{4x}  = L_{4x}\big|_{b_{k=1}} \ , \qquad 
L_{4x}  \te = -\frac{1}{8}b_1 (\partial \bm x\cdot\partial \bm x)^{2}
+\frac{1}{4}b_2 (\partial x^{i}\partial x^{j})(\partial x^{i}\partial x^{j})
-\frac{1}{4}b_3\bm x^{2}(\partial \bm x\cdot \partial \bm x)-\frac{1}{2}b_4 (\bm x^{2})^{2}  \ . 
\ea
Then using the dimensional regularization relations \rf{333}-\rf{335} we  find
\ba
\la{5.5}
\langle  L_{4x}  \rangle=  q_\rmb\, \gxx^{2}\,,   \quad \te 
q_\rmb = - b_{1}\,\frac{N_{x}(2+(d+1)N_{x})}{2(d+1)}+ b_{2}\frac{N_{x}(2+d+N_{x})}{d+1}+ 
\frac{1}{2}b_3 N_{x}^{2} - \frac{1}{2} b_4 N_{x}(N_{x}+2)\,,  \quad N_x=3\, , 
\ea
where $N_x$ stands for the number of $x^i$ fields.\foot{Note that in the 
 dimensional reduction regularization  the derivatives of scalar 
  bosons   are treated as  in  the 
standard  dimensional regularization, i.e. they are  assumed  to be \rd=$2-2\eps$ dimensional ones.  If their  indices were  directly 
contracted in 2  instead of $2-2\eps$  dimensions 
that would    be equivalent to  setting  $d=1$ in  the coefficient in \rf{5.5}  and then the  final  result in \rf{449} would be zero.} 

Expanding the coefficient $q_\rmb $  for $d=1-2 \ep\to 1$   we get  
\be\la{56} 
q_\rmb\big|_{d\to 1} = \tfrac{1}{2}N_{x}\big[-b_{1}(1+N_{x})+b_{2}(3+N_{x})+b_{3}N_{x}-b_{4}(2+N_{x})\big]  
+  \tfrac{1}{2}N_x \big[-b_{1}+b_{2}(N_{x}+1)\big] \ep  +\mc O(\ep^{2}).
\ee
Since according to \rf{4.21}  $\gxx^2= ( {1\ov 4 \pi \ep} + ...)^2 $
we conclude   that the double pole term in \rf{5.5} cancels out   for the relevant values of 
$b_k=1$ for {\it any}  value of $N_x$.
The coefficient of the remaining  single pole is then   proportional to 
$  \tfrac{1}{2}N_x \big[-b_{1}+b_{2}(N_{x}+1)\big] = \ha N_x^2 $. 

We conclude  that in the case  of  $b_k=1$  the bosonic contribution to $f_2$  in  \rf{319}  is given   by 
\be 
f_{2\, \rm b}  = \langle \mc L_{4x}\rangle = q_{\rmb}\,  \gxx^2\ , \qquad \ \ \ \ \ \ 
q_{\rmb}\big|_{b_k=1}=  {N_x^2(1-d)\ov 2(1+d) } =  {9(1-d)\ov 2(1+d) } \ .   \la{449}
\ee

\subsection{Fermionic contribution \la{4.2s}}

In computing the expectation values of the  fermionic terms 
  we will assume the formal continuation to the  2d Euclidean signature in the GS action   and thus in the  fermionic  propagator
 $ \langle\theta(\ww)\bar{\theta}(\ww') \rangle = S(\ww, \ww')$ (cf. \rf{4.39}). 
The  fermionic contributions  to $f_2$  are given by 
\be\la{501}
f_{2\, \rm bf} = i  \langle  \L_{x^2\theta^2} \rangle \ , \qquad \qquad \ \ 
 f_{2\, \rm f} =  \langle  \L_{\theta^4} \rangle \ ,  \ee
where 
the relevant terms in the fermionic part of the GS Lagrangian in \rf{2431} 
 may be  written as\foot{
 We follow the notation  in  section \ref{s2f},  i.e.  $\hGamma = \G_{04}, \ \Gamma_* =  i \Gamma_{01234} , \ \Gamma'_* =  i \Gamma_{56789}$, $\Gamma_{11} = \Gamma_* \Gamma'_*$, $\g_\a $ is the \adst  2-bein   projection of $\G_{\halpha}=(\G_0,\G_4)$,  etc.} 
\ba
 \L_{\rm f} = & \L_{\theta^2} +  \L_{x^2\theta^2}  + \L_{\theta^4} \ ,
    \qquad \ \  \L_{x^2 \theta^2 } =   L_{x^2 \theta^2 } \big|_{c_1=1} \ , \qquad   \L_{ \theta^4 } 
    =   L_{\theta^4 } \big|_{c_k=1}  \la{151} \\
\L_{\theta^2} = & \tfrac{i}{2}\bar\theta\slashed{\mk D}\theta  \ , \ \ \ \qquad  \ \    \mk D_{\alpha} = \nabla_{\alpha}+\tfrac{1}{2}\g_{\alpha}\G_{123}, \qquad \slashed{\mk D} = \slashed{\nabla}+\G_{123}
 \ , \la{152} \\
 \te L_{x^2 \theta^2 } = &\te \frac{i}{4}(\bm{x}^{2}+\frac{1}{2}\partial_{\a}\bm{x}\cdot\partial^{\a}\bm{x})\ \bar\theta\slashed{\mk D}\theta
+\frac{i}{4}(\bm{x}^{2}+\frac{1}{2}\partial_{\a}\bm{x}\cdot\partial^{\a}\bm{x})\ \bar\theta\G_{123}\theta 
-\frac{i}{4} c_1 \partial^{\alpha}\bm{x}\cdot\partial^{\beta}\bm{x}\ \bar\theta\g_{\alpha}\mk D_{\beta}\theta
\ , \la{153} \\
L_{\theta^4} = &\te  \,\frac{1}{96}c_2 \bar\theta\G_{123}\theta\ \bar\theta\slashed{\mk D}\theta
+\frac{i}{192}c_3\bar\theta\g^{\alpha}\G^{ij}\theta\ \bar\theta\hG\G_{ij}\G_{*}\mk D_{\alpha}\theta
-\frac{i}{192}c_4 \bar\theta\g^{\alpha}\G^{a'b'}\theta\ \bar\theta\hG\G_{a'b'}\G_{*}'\mk D_{\alpha}\theta\lp\te 
+\frac{1}{16} {c_{5}}\,(\bar\theta\slashed{\mk D}\theta)^{2}- {1\ov 16} {c_{6}}\,\bar\theta\g_{\alpha}\mk D_{\beta}\theta\ \bar\theta\g^{\beta}\mk D^{\alpha}\theta \ .  \la{154} 
\ea
We have omitted   mixed boson-fermion terms   with the massless  field $\del y \del y$  factors 
  as they will not   contribute   to the 2-loop free energy  (cf. \rf{310}).
As  in \rf{041}  we  introduced  some auxiliary  coefficients $c_1, ..., c_6$   (equal  to 1 in the GS Lagrangian in \rf{2431})  as 
 this  will allow us  to trace the contributions of the  individual   quartic vertices. 

Let us  first  observe that the first two terms in \rf{153}   will not contribute  to \rf{501} 
  since  due to \rf{310}  and \rf{334} (or, equivalently, the    $m^2=2$   scalar equation of motion) we have 
\be
\la{5.10}
\langle\partial_{\alpha}\bm{x}\cdot\partial_{\beta}\bm{x} \rangle= -\tfrac{2}{d+1}g_{\alpha\beta}\langle\bm{x}^{2}\rangle\ ,\qquad  \qquad 
\langle\partial_{\a}\bm{x}\cdot\partial^{\a}\bm{x}\rangle = -2\langle \bm{x}^{2}\rangle.
\ee
As already   mentioned, we will use the  dimensional reduction prescription, i.e. will keep $\G_\halpha=\gamma_\halpha$  in 2 dimensions, so that 
\ba
\la{5.11}
 \{\g_{\alpha}, \g_{\beta}\} = 2\bar g_{\alpha\beta}, \qquad \bar g^{\a\b}  \bar g_{\alpha\b }=2\  ,\qquad  g^{\alpha\beta}\bar g_{\alpha\beta} = 2\ , 
\ea
where $\bar g_{\a\beta}$ is   the metric of \adst. Also, contracting $\g_\a$ with derivatives   will 
select only the 2d  components of the latter. 
 
 Note that under $ \langle ... \rangle $ one may  use \rf{5.10}  to rewrite  the  contribution of the 
 remaining $c_1 $ term in \rf{153}  using \rf{152} as 
\be 
\langle L_{x^2 \theta^2 } \rangle = 
-\tfrac{i}{4} c_1 \langle \partial^{\alpha}\bm{x}\cdot\partial^{\beta}\bm{x}\ \bar\theta\g_{\alpha}\mk D_{\beta}\theta \rangle 
=   \tfrac{i}{2(d+1)} c_1\langle  \bm{x}^{2}\bar\theta  \slashed \DD  
\theta \rangle \ . \la{566}
\ee
This term is  thus proportional to the expectation value  of the 
fermion kinetic term  suggesting   that its contribution should be scheme-dependent 
(formally, it can be redefined away but it  will still contribute  under the dimensional reduction regularization, cf. \rf{33710}).

Using the definition of $\DD_\a$ and the  Majorana   spinor relation
$
\bar\theta\g_{\alpha}\g_{\beta}\G_{123}\theta = \bar g_{\alpha\beta}\bar\theta\G_{123}\theta $ 
we  may rewrite \rf{154} as 
\ba
L_{\theta^4} = &\te 
\frac{1}{96} (c_2 + 12 {c_{5}}- 6 {c_{6}})\bar\theta\G_{123}\theta\ \bar\theta\slashed{\nabla}\theta
+ \frac{1}{96} (c_2 + 6 {c_{5}}- 3 {c_{6}})(\bar\theta\G_{123}\theta)^{2}\lp \te 
+\frac{i}{192}c_3 \bar\theta\g^{\alpha}\G^{ij}\theta\ \bar\theta\hG\G_{ij}\G_{*}\nabla_{\alpha}\theta
- \frac{i}{192}c_4\bar\theta\g^{\alpha}\G^{a'b'}\theta\ \bar\theta\hG\G_{a'b'}\G_{*}' \nabla_{\alpha}\theta\lp\te 
+\frac{i}{384}c_3 \bar\theta\g^{\alpha}\G^{ij}\theta\ \bar\theta\hG\G_{ij}\G_{*}\g_{\alpha}\G_{123}\theta
- \frac{i}{384}c_4 \bar\theta\g^{\alpha}\G^{a'b'}\theta\ \bar\theta\hG\G_{a'b'}\G_{*}' \g_{\alpha}\G_{123}\theta\lp\te 
+\frac{1}{16}c_{5}(\bar\theta\slashed{\nabla}\theta)^{2}
-\frac{1}{16} c_{6}\bar\theta\g_{\alpha}\nabla_{\beta}\theta\ \bar\theta\g^{\beta}\nabla^{\alpha}\theta
\ .\la{4133}
\ea
Using also  that $\G_{11}\theta=\theta$   and ($i,j=1,2,3; \ a'=5,...,9$)
\ba\la{4145}
 \hG\G_{ij}\G_{*} = -i\G_{ij}\G_{123},\ \qquad \ \ \ 
 \hG\G_{a'b'}\G_{*}' =  \hG\G_{a'b'}\G_{*}\G_{11}=-i\G_{a'b'}\G_{123}\G_{11}\ , 
\ea
we may also rewrite \rf{4133} as 
\ba
L_{\theta^4} = &\te 
\frac{1}{96} (c_2 + 12 {c_{5}}- 6 {c_{6}})\bar\theta\G_{123}\theta\ \bar\theta\slashed{\nabla}\theta
+ \frac{1}{96} (c_2 + 6 {c_{5}}- 3 {c_{6}})(\bar\theta\G_{123}\theta)^{2}\lp \te 
+\frac{1}{192}c_3 \bar\theta\g^{\alpha}\G^{ij}\theta\ \bar\theta\G_{ij}\G_{123}\nabla_{\alpha}\theta
- \frac{1}{192}c_4\bar\theta\g^{\alpha}\G^{a'b'}\theta\ \bar\theta\G_{a'b'}\G_{123}\nabla_{\alpha}\theta\lp\te 
+\frac{1}{384}c_3 \bar\theta\g^{\alpha}\G^{ij}\theta\ \bar\theta\G_{ij}\g_{\alpha}\theta
- \frac{1}{384}c_4 \bar\theta\g^{\alpha}\G^{a'b'}\theta\ \bar\theta\G_{a'b'}\g_{\alpha}\theta\lp\te 
+\frac{1}{16}c_{5}(\bar\theta\slashed{\nabla}\theta)^{2}
-\frac{1}{16} c_{6}\bar\theta\g_{\alpha}\nabla_{\beta}\theta\ \bar\theta\g^{\beta}\nabla^{\alpha}\theta
\ .\la{41338}
\ea
To compute the expectation values of \rf{566} and \rf{4133}  we need   to take into account 
that  $\theta$ are Majorana.\foot{The propagator   corresponds  to the quadratic action  factor in the path  integral with 
$\exp{( i S_M)}  \to \exp[ -  \ha \int \bar \theta  (\slashed \nabla + \G_{123})   \theta ]$ so that $\langle  \theta \bar \theta \rangle = (\slashed \nabla + \G_{123})^{-1}  $ 
(see  also Appendix \ref{apA}).  
The  Majorana action is $\frac{1}{2}\int \bar\theta K\theta= \frac{1}{2}\theta^T C K\theta $
(with $CK$ operator  being antisymmetric)  so that  the propagator should be same as that of 
the Dirac action $\int \bar\theta K \theta$.}
As a result, we  get  switching to the  Euclidean propagators  in $\langle ... \rangle$  
(using \rf{333},\rf{3344},\rf{33710},\rf{ba1}  and that $\theta$  are subject to \rf{229})\foot{A check
 of  normalization: 
we have  from  \rf{152} and \rf{566} 
 $ i L=  \tfrac{i}{2} [1 +    { c_1 \ov d+1 }   \bm{x}^{2}]  \bar\theta  \slashed \DD  
\theta + ... $.  Then integrating over $\theta$   the free energy   contribution  of this term will  be   given by 
$F = - \ha \tr \log ( [1 +    { c_1 \ov d+1 }   < \bm{x}^{2}> ]    \slashed \DD  ) 
= - \ha  { c_1 \ov d+1 } < \bm{x}^{2}> \nf  \delta(0) $  where $\nf=16$ and  the  fermionic $\delta(0)$ is given by $ \delta(0) = \te { d-1 \ov d+1}  \sss  $
in dimensional  reduction regularization.}
\ba
\la{525}
& 
f_{2\, \rm bf}=i \langle L_{x^2\theta^2} \rangle =  - \te {c_1\ov 2(d+1)} \,\langle\bm{x}^{2} \rangle\  J = - \te {c_1\ov 2(d+1)}  N_x \gxx\, J  \ , \ \ \ \ \ \   J= 
 \langle \bar\theta(\slashed{\nabla}+\G_{123})\theta\rangle=
J_1 + J_2   \ , \\
&J_{1} = \langle \bar\theta\slashed \nabla   \theta    \rangle = -  \te {2\ov d+1} \nf  \, \sss  
\ ,   \qquad 
J_2 =\ \langle \bar\theta\G_{123}  \theta    \rangle  =\nf \,  \sss\  , \qquad \ \  J  = \te { d-1 \ov d+1}  \nf \, \sss \ , 
  \la{75}  \\ 
& \qquad \qquad N_x=3 \ ,  \qquad \ \ \  \nf \equiv \tr P_+ =16 \ . \la{76}
\ea 
Similarly,      the  contribution of the quartic fermionic  terms in \rf{4133} 
  is given by   (see  Appendix \ref{apA}) 
\ba
& \ \ \qquad  \qquad \qquad f_{2\, \rm f} = \langle L_{\theta^4} \rangle = \sum_{k=1}^8 I_k \ , \la{800}\\
I_{1} &= \tfrac{1}{96}(c_{2}+12 c_{5}-6c_{6}) \bi_{1} \ , \qquad  \qquad 
\bi_{1}  = \langle\bar\theta\slashed{\nabla}\theta\ \bar\theta\G_{123}\theta\rangle 
= - \tfrac{2}{d+1} \nf ( \nf -2) \, \sss^2 \ ,\la{77} \\
I_{2} &=  \tfrac{1}{96} ({c_{2}+6c_{5}-3c_{6}})\bi_{2} \ , \qquad \qquad  \bi_{2}
=      \langle \bar\theta\G_{123}\theta\,  \bar\theta\G_{123}\theta\rangle =   \nf (\nf -2) \sss^2 \ , \la{78}    \\ 
I_{3} &= \tfrac{1}{192} c_3   \bi_{3} \ , \qquad  \qquad  \bi_{3} =    \langle\bar\theta\g^{\alpha}\G^{ij}\theta\ \bar\theta\G_{ij}\G_{123}\nabla_{\alpha}\theta\rangle = - \te {24 \ov d+1}  \nf \sss^2      \,,\la{79} 
 \\
I_{4} &= - \tfrac{1}{192} c_{4}\bi_{4}  \ , \qquad  \qquad 
\bi_{4}= \langle\bar\theta\g^{\alpha}\G^{a'b'}\theta\ \bar\theta\G_{a'b'}\G_{123} \nabla_{\alpha}\theta\rangle = 
- \te {80 \ov d+1} \nf  \sss^2 \ ,\la{80} \\
I_{5} &=\tfrac{1}{384}c_3 \, \bi_{5} \ , \qquad \qquad  \bi_{5} =
   \langle\bar\theta\g^{\alpha}\G^{ij}\theta\ \bar\theta\G_{ij}\g_{\alpha}\theta\rangle =   
24 \nf \sss^2 \ , \la{81} \\
I_{6} &=-\tfrac{1}{384}c_4 \,  \bi_{6} \ , \qquad  \qquad  \bi_6 =
\langle\bar\theta\g^{\alpha}\G^{a'b'}\theta\ \bar\theta\G_{a'b'} \g_{\alpha}\theta\rangle
 =80 \nf \sss^2 \ , \la{82} \\
I_{7} &= \tfrac{1}{16} c_5 \, \bi_7, \qquad \qquad  \bi_7=\langle  \bar\theta\slashed{\nabla}\theta\,  \bar\theta\slashed{\nabla}\theta \rangle=
\Big[ \tfrac{4}{(d+1)^2}  (\nf-1) - \ha + \tfrac{d (d+1) -4 }{ 2 (d+1) } \Big] \nf\, \sss^2 \ , \la{83} \\
I_{8} &=- \tfrac{1}{16} c_6 \, \bi_{8}, \qquad  \qquad \bi_{8}
=\langle  \bar\theta\G_\a {\nabla}_\b\theta\,  \bar\theta\G^\b {\nabla}^\a \theta \rangle=
\Big[ \tfrac{2}{(d+1)^2}  (\nf-2) + \ha + \tfrac{d (d+1) -4 }{ 2 (d+1) } \Big] \nf\, \sss^2 \ . \la{84} 
\ea
As a result, we get for \rf{525},\rf{800}  
\ba
\la{535}
&\qquad \qquad f_{2\, \rm bf} =q_\rmbf \, \gxx\,\sss\ , \qquad \,  
\qquad   q_\rmbf  = \,c_{1}\, \frac{24(1-d)}{(d+1)^{2}} \ , 
\\
&\qquad \qquad  \la{533} 
f_{2\, \rm f} =  q_\rmf \, \sss^{2}\ , \qquad \ \ \    \ \qquad  q_\rmf =-{\rm c}\,  \frac{1-d}{6\,(d+1)^{2}}\,   , \qquad \qquad 
\ \ \  {\rm c} \big|_{c_k=1}=  -12 (5-3d)\\
&{\rm c}  =4 (7 c_2+3 c_3-10 c_4- 6c_6)
+2 (7 c_2+3 c_3-10 c_4+48 c_5- 30 c_6) (d-1)
 +3 (c_5-c_6) (d-1)^2
  \ . \la{5388}
\ea
Setting all $c_k=1$  we thus   find the values of $q_\rmbf $ and $q_\rmf $  given in \rf{17}.

Note that the contributions 
of the $c_2,c_3$ and $c_4$ terms  in  \rf{154} effectively cancel  each  other in $\rm c$, 
i.e. $f_{2\, \rm f}$ receives non-zero contributions only from the $c_5$ and $c_6$ terms. 
 Also, the  coefficient of the   simple pole  in 
$f_{2\, \rm f}$   (coming only from the  first  $4 (7 c_2+3 c_3-10 c_4- 6c_6)$   term in  $\rm c$ in \rf{5388}) 
 does not depend on  the contribution  of the $c_5 (\bar\theta\slashed{\mk D}\theta)^{2}$ term in \rf{154}.\foot{This  is related to the fact that in the dimensional reduction regularization $\langle \bar\theta\slashed{\mk D}\theta \rangle = \OO (1)$, cf. \rf{33710},\rf{75}.}

To conclude,  the total    value of the 2-loop coefficient is  found to be 
 \ba
\la{5.41}
f_2 = q_\rmb \gxx^2 + q\rmbf \gxx \sss + q_\rmf \sss^2 =
    \frac{9(1-d)}{2(1+d)}\,\gxx^{2}  +  \frac{24(1-d) }{(1+d)^{2}}\, \gxx\, \sss + \,\frac{2(1-d)(5-3d)}{(1+d)^{2}}\,\sss^{2} \ , 
\ea
where $\gxx$ and $\sss$ are given by  \rf{4.21}  and \rf{3344} (or \rf{1.8} and \rf{9}). 


\section{Concluding remarks}

As was already discussed in the Introduction  (see \rf{101}), expanding  \rf{5.41} in $\ep= \ha (1-d)\to 0$  we find that 
the pole  term does not cancel   while the finite remainder  depends on the constant parts in the  bosonic and fermionic propagators in \rf{1.8},\rf{9}, i.e.  
\be\la{202} 
f_2 = -\frac{11}{32\pi^{2}\, \eptt} + f_{2,\rm fin}  \ , \qquad \qquad 
f_{2,\rm fin} =   {6 a_\rmb  + 16 a_\rmf  -19 \ov 32\pi^2}  \ , \qquad \ \ \  f_{2,\rm fin}\big|_{a_\rmb =2, \,  a_\rmf =1} =  \frac{9}{32\pi^{2}}  \ . 
\ee
As in the S-matrix computation in flat target space \ci{Beccaria:2025xry}  it is natural to suggest that 
the pole term should be  subtracted or cancelled  by adding a particular 2-loop  counterterm 
 that    should be   required  by underlying symmetries of the theory (supersymmetry   and integrability). 
 The regularization or  scheme choice   consistent with  symmetries    should then lead to the finite part of $f_2$ that should match the 
 dual gauge theory prediction in \rf{4}. 
 
Clearly, more is to be done to substantiate this suggestion. In particular, it would be important  to clarify if the dimensional reduction regularization that  we used  actually  preserves  the residual \adst  supersymmetry   that remains in the $\kappa$-symmetry gauge \rf{255}. 

In general, one  may  use also other regularizations like $\zeta$-function or heat kernel one  applied  directly 
in  2 dimensions
(see Appendix  \ref{apB}). In this case  the $\delta^{(2)}(w,w)$  or ``$\delta(0)$''  terms  may  have a non-zero 
finite part. 
  As a result, one may get additional 2-loop   contributions
from local path integral measure
(cf. \ci{Andreev:1990iv})  and  determinant of  
 local $\kappa$-symmetry ghost  operator corresponding to the gauge \rf{255}.
 In general,   when  such   ``$\delta(0)$''  terms  are non-zero, 
the result  may depend  on  field redefinitions and gauge choices 
unless one carefully takes into account all measure and algebraic  ghost
factors.\foot{In fact, 
 these  reservations apply also to  our treatment of fermions in dimensional reduction regularization
 where   the fermionic   $\delta^{(2)}(w,w)$  is no longer zero:  the equation of motion
   for the propagator  is satisfied modulo $\OO(\eps)$ term,  see  \rf{33710}.}

To  try to fix the regularization/subtraction scheme  ambiguities it would  be important to compute string   world-sheet 
loop corrections to  other 
related observables (like   latitude WL  or correlators of operators on   an  infinite WL)  and compare them to
  data on the dual gauge theory side. 
  A  simple example of 1-loop correction to 2-point function of $x^i$ fluctuations   is discussed  in  Appendix \ref{apC}.

It would be  interesting   also to repeat a similar 2-loop  computation in the case of $\ha$ BPS  WL in 
ABJM  theory which in  the planar limit   is dual  to  IIA string in AdS$_{4}\times \CP^{3}$ (the corresponding setup is reviewed
in Appendix \ref{apD}).


\iffa 
add about   computing ratio to 1/4 BPS WL -- 
comments from message to Zarembo. But different induced metric so divergences need not cancel in the ratio. 

\fi

\iffa   \foot{Previous expectation  was that GS string  result is log UV finite;
there could be still scheme dependence -- depending  e.g. on  how  one
deals   with possible  power div, measure etc     but hope   was  that
there should be a scheme in which result matches  CFT.
If there are genuine 2-loop divergences I do not see how  comparing
circular loop to  some other WL  may help --
if divergence depends on surface  both results will be scheme dependent.
The point is that if both results are finite, then to eliminate the
ambiguity if possible finite scheme choice (redefinition of tension)
one  is to express one result in terms of the other   to eliminate
dependence on tension.
In general, if string side  is not UV-defined AdS/CFT is in trouble --
 for example, we assume that  inverse tension  corrections to
anomalous dimensions  like Konishi
 can be  computed systematically and matched to integrability-based
(assumed)    approach.
It may of course be that GS string requires particular completion by
higher derivative counterterms that cancel that divergences but at the
moment it is not clear how to fix them.
We had an indication of that
in 2-loop S-matrix in flat space  \ci{Beccaria:2025xry}
 that is not manifestly finite  but given that dim reg is problematic
for GS string that could be  an excuse. 
 }\fi



\section*{Acknowledgements}
We thank  S. Giombi, R. Metsaev  and  J. van Muiden    for useful   discussions  and   comments on the draft.
 MB is supported by the INFN grant GAST. 
 SAK acknowledges the support of the President's PhD Scholarship of Imperial College London.
The work of AAT   is supported by the STFC grant ST/T000791/1.



\appendix

\section{Correlators of Majorana fermions}
\la{apA}
Our  conventions   for the 10d MW    spinors were discussed in \rf{221}--\rf{215}. 
Useful identities are, e.g., 
\be
\la{A.8}
\bar\theta\Gamma_{A}\theta=0, \qquad 
\bar\theta\Gamma_{ABC}\partial\theta = \partial\bar\theta\Gamma_{ABC}\theta\to \bar\theta\Gamma_{ABC}\partial\theta=\tfrac{1}{2}\partial(\bar\theta\Gamma_{ABC}\theta).
\ee
We need to  compute  the expectation values of the  fermion operators of the form 
$
\langle\bar\theta   X \theta\rangle $ and $ \langle\bar\theta  X \nabla_{\alpha}\theta\rangle$
where $X$ is a product  of  $\G_A$  matrices (with $(C X)^{T} = -CX$)  and  the fermion propagator is defined as 
\be\la{ba1}
\langle \theta_u(\w)\bar\theta_v(\w') \rangle = S_{uv}(\w,\w')\  ,  \qquad 
\langle \theta_u \theta_v  \rangle = (S\, C^{-1})_{uv}\  , \qquad (SC^{-1})^{T} = -SC^{-1} \ . 
\ee
Here  $u,v$ are 10d spinor indices  and the chirality projector $P_+=\frac{1}{2}(1+\Gamma_{11})$ is implicit. 
Then
\be
\langle\bar\theta  X \theta\rangle = 
\tr[SC^{-1}(CX)^{T}] = -\tr[SX], \qquad \ \ \langle\bar\theta X
 \theta'\rangle  = -\tr[S' X] \ , 
\ee
where prime stands for a derivative  over  the second argument. 

If $X$ and $\wt X$ are  products of $\G_A$ matrices then one finds that 
\ba
\la{B.12}
 \langle \bar\theta X\theta\ \bar\theta\wt X\theta\rangle 
= \tr[X S]\tr[\wt X S]-2\tr[XS\wt X S].
\ea
In the case   when $\theta'$ and $\theta''$  stand for derivatives of $\theta$ 
  we get 
\ba
& \langle \bar\theta X\theta'\ \bar\theta\wt  X\theta''\rangle 
= \tr[X S']\tr[\wt X S'']-\tr[X S'\wt X S''] {-\tr[X \hat S C^{-1}\wt X^{T}C S]}, \la{b12} \\
&\langle\theta' \theta \rangle =  S'C^{-1} , \qquad
\langle\theta'' \theta \rangle =  S''C^{-1} , \qquad
\langle\theta' \theta'' \rangle = \hat S C^{-1} . \la{b13}
\ea
In particular, 
\be
\la{B.18}
\langle(\nabla_{\alpha}\theta(\w))_{u}\theta_{v}(\w')\rangle  = (\nabla_{\alpha}SC^{-1})_{uv} \ , \quad 
\langle(\nabla_{\alpha}\theta)_{u}\ (\nabla_{\beta}\theta)_{v}\rangle = (\hat S_{\alpha\beta}C^{-1})_{uv}\  , \quad 
\nabla_{\alpha}=\partial_{\alpha}+\frac{1}{2\sigma}\gamma_{\hat 0\alpha}.
\ee
Using \rf{4.39}--\rf{3344}   we  find that 
\be
\la{B.23}
\hat S_{\alpha\beta}(\w,\w) 
 = -\frac{1}{4z^{2}}\Big[\G_{\halpha\hbeta}+\frac{4m^2 -d(d+1)}{d+1}\eta_{\halpha\hbeta}\Big] \Gamma_{123}\sss\ . 
\ee
To compute the  traces  over the spinor indices we note that  after the $\kappa$-symmetry gauge  fixing \rf{255}   the remaining 
 $D=10$ Majorana-Weyl  fermion $\theta$  in \rf{229}  has  $N_\theta =16$ independent real components 
   ($\tr I =32, \ \tr \Gamma_{11}=0$)  so that 
\ba
 &\qquad  \tr P_{+}= N_\theta=16, \qquad\qquad   P_+ =\ha (1 + \G_{11}) \ , \la{a9}\\
&\tr(P_{+}\G_{A}\G_{B}) =\eta_{AB}\, \tr P_{+}, \qquad 
\tr(P_{+}\G_{A}\G_{B}\G_{C}\G_{D}) = (\eta_{AB}\eta_{CD}-\eta_{AC}\eta_{BD}+\eta_{AD}\eta_{BC})\, \tr P_{+}\ . \no 
\ea
Then we  find  for the correlators in  \rf{75}
\ba
\la{C.2}
&J_1=\langle\bar\theta\slashed{\nabla}\theta\rangle=-\tr[\g^{\alpha}P_{+}\nabla_{\alpha}S] 
=-\frac{m\,}{d+1}  \sss \tr[P_{+}\g^{\alpha}\g_{\alpha}] = -\frac{2\,m\, }{d+1} \sss \, \nt \ , 
\\ 
\la{C.1}
&J_2=\langle\bar\theta\G_{123}\theta\rangle=-\tr[\G_{123}P_{+}S] =  \sss\, \nt \ .
\ea
Similarly,  we reproduce the  expressions for the 4-fermion  correlators in \rf{77}--\rf{84}\foot{To recall, $\g_{\hat \a} = \G_{\hat \a}$   where $\hat \a$ is the tangent-space 2d index with (0,4) values.}
\ba
\bi_{1} &= \langle\bar\theta\slashed{\nabla}\theta\theta\ \bar\theta\G_{123}\theta\rangle = 
\tr[P_{+}\g^{\alpha}\nabla_{\alpha}S]\,\tr[P_{+}\G_{123}S]-2\tr[P_{+}\g^{\alpha}\nabla_{\alpha}S\G_{123}S]
= -\frac{2 m\,}{d+1}\sss^{2} \nt(\nt-2), \no  \\
\bi_{2} &= \langle\bar\theta\G_{123}\theta\theta\ \bar\theta\G_{123}\theta\rangle = 
\tr[P_{+}\G_{123}S]\,\tr[P_{+}\G_{123}S]-2\tr[P_{+}\G_{123}S\G_{123}S]
= \sss^{2}  \nt(\nt-2), \no  \\
\bi_{3} &= \langle \bar\theta\G_{ij}\G_{123}\nabla_{\alpha}\theta\ \bar\theta\g^{\alpha}\G^{ij}\theta\rangle = 
\tr[P_{+}\G_{ij}\G_{123}\nabla_{\alpha}S]\,\tr[P_{+}\g^{\alpha}\G^{ij}S]-2\tr[P_{+}\G_{ij}\G_{123}\nabla_{\alpha}S\g^{\alpha}\G^{ij}S]\lp
= \frac{m\,\sss^{2}}{d+1}\Big(
\tr[P_{+}\G_{ij}\g_{\alpha}\G_{123}]\,\tr[P_{+}\g^{\alpha}\G^{ij}\G_{123}]-2\tr[P_{+}\G_{ij}\G_{123}\g_{\alpha}\g^{\alpha}\G^{ij}\G_{123}]
\Big)
 = -24\frac{m\,}{d+1}\sss^{2} \nt , \no  \\  
\bi_{4} &= \langle \bar\theta\G_{a'b'}\G_{123}\nabla_{\alpha}\theta\ \bar\theta\g^{\alpha}\G^{a'b'}\theta\rangle  
= 4\, \frac{m\,}{d+1}\sss^{2} \tr[P_{+}\G_{a'b'}\G^{a'b'}] = -80\,\frac{m\,}{d+1}\sss^{2} \nt \ , \la{a12}  \\
\bi_{5}  &= \langle \bar\theta\g_{\alpha}\G_{ij}\theta\ \bar\theta\g^{\alpha}\G^{ij}\theta\rangle
=  \sss^{2}\Big(\tr[P_{+}\g_{\alpha}\G_{ij}\G_{123}]\tr[P_{+}\g^{\alpha}\G^{ij}\G_{123}]
-2\tr[P_{+}\g_{\alpha}\G_{ij}\G_{123}\g^{\alpha}\G^{ij}\G_{123}\Big)\no \\
&=24\,\sss^{2}\,\nt ,\no  \\
\bar I_6 &= \langle\bar\theta\g^{\alpha}\G^{a'b'}\theta\ \bar\theta\G_{a'b'}\g_{\alpha}\theta\rangle  
= \sss^{2}\,\Big(
\tr[P_{+}\g^{\alpha}\G^{a'b'}\G_{123}]\tr[P_{+}\g_{\alpha}\G_{a'b'}\G_{123}]
-4\tr[P_{+}\G^{a'b'}\ \G_{a'b'}]\Big)\no \\
&= 80\,\,\sss^{2}\,  \nt \ , \no
\ea 
 where 
$
\tr[P_{+}\g_{\alpha}\G_{a'b'}\G_{123}] = 0,\ 
\tr[P_{+}\G^{a'b'} \G_{a'b'}] = -20\,  \nt$. 
Also, 
\ba
\la{C.7}
\bar I_7  = & \langle\bar\theta\slashed{\nabla}\theta\ \bar\theta\slashed{\nabla}\theta\rangle  
\lp
= \tr[\g^{\alpha}P_{+}\nabla_{\alpha}S]\tr[\g^{\beta}P_{+}\nabla_{\beta}S]
-\tr[\g^{\alpha}P_{+}\nabla_{\alpha}S\ \g^{\beta}P_{+}\nabla_{\beta}S]
-\tr[\g^{\alpha}P_{+}\hat S_{\alpha\beta}C^{-1}(\g^{\beta})^{T}C\, P_{+}S]
\lp
=4\,\big(\frac{m\,}{d+1}\big)^{2}\,\sss^2  \nt (\nt-1) 
+\sss^{2}\Big[-\ha +\frac{d(d+1)-4m^{2}}{2(1+d)}\Big]\, \nt \ , 
\\
\la{C.14}
 \bi_8 = &\langle  \bar\theta\g_{\alpha}\nabla_{\beta}\theta\  \bar\theta\g^{\beta}\nabla^{\alpha}\theta\rangle 
\lp
= 
\tr[ \g_{\a}P_{+}\nabla_{\beta}S]\tr[\g^{\beta}P_{+}\nabla^{\alpha}S]
-\tr[ \g_{\a}P_{+}\nabla_{\beta}S\ \g^{\beta}P_{+}\nabla^{\alpha}S]
-\tr[ \g^{\a}P_{+}\hat S_{\beta\alpha}C^{-1}(\g^{\beta})^{T}C\,P_{+}S]
\lp
= 2\big(\frac{m\,}{d+1}\big)^{2}\sss^2 \nt (\nt-2) +\sss^{2}\,\Big[
\ha +\frac{d(d+1)-4m^{2}}{2(d+1)}
\Big]\nt,
\ea
  where we used that 
$
\g^{\alpha}\g_{\alpha\beta}\g^{\beta} =2$  and  \rf{B.23}.

\section{General  form of   2-loop correction   in 2d   and $\zeta$-function regularization  \la{apB}}

Here  we shall not use  dimensional regularization, i.e. 
 assume   that  we regularize the theory  directly in 2 dimensions. In particular,   we may 
 adopt  the $\zeta$-function regularization. 
 
We shall    use   the following notation ($\xi^\a$  are coordinates  in  \adst  space)
\begin{align}
    &\langle x^{i}(\xi)x^{j}(\xi') \rangle = \delta^{ij}G_{x}(\xi, \xi') \ , \quad \langle y^{a'}(\xi)y^{b'}(\xi') \rangle = \delta^{a'b'}G_{y}(\xi, \xi') \ ,\\
    &    \langle\theta(\xi)\bar{\theta}(\xi') \rangle = G_{\theta}(\xi, \xi') \ , \qquad \qquad   i \slashed{\mathfrak{D}}G_{\theta } =\ri   \delta^{(2)} \ . 
    \la{b1}
\end{align}
In the coincident-point   limit we define\foot{In this Appendix 
  we define the fermionic  Green's function $G_{\theta}(\xi, \xi')$ 
 corresponding  to the quadratic Lagrangian in  \rf{404}
so that there is an extra factor of $i$ compared  to the  definition   of $S$ in \rf{3166}, \rf{4.33}, i.e. 
$G_{\theta}= - i S$. The  constant $\rm S$  in \rf{3344} is then related to  the constant 
$\rm G_{\theta}$  
defined  in \rf{bb2}  as  $\rm G_{\theta}=   \rm S$.}   
\begin{align}
&\lim_{\xi'\rightarrow\xi}G_{x, y}(\xi, \xi') =  \rfe_{x, y} \ , \qquad\qquad  \lim_{\xi'\rightarrow\xi}\partial_{\alpha} G_{x, y}(\xi, \xi')=\lim_{\xi'\rightarrow\xi}\partial'_{\alpha} G_{x, y}(\xi, \xi') = 0 \ ,  \la{bb1} \\
    &\lim_{\xi'\rightarrow\xi}\partial_{\alpha}\partial'_{\beta} G_{x, y}(\xi, \xi') =  \rh_{x, y} \, g_{\alpha\beta} \ , \label{b2} \\
    &\lim_{\xi'\rightarrow \xi} G_{\theta}(\xi, \xi') =  \rfe_\theta   M \ , \qquad \lim_{\xi'\rightarrow \xi} i\mathfrak{D}_{\alpha}G_{\theta}(\xi, \xi') = \rh_\theta  \gamma_{\alpha} \ ,  \ \ \ \ \   M= - i \G_{123} \ , \la{bb2}  \\
    &\lim_{\xi'\rightarrow \xi}\mathfrak{D}_{\alpha}\mathfrak{D}'_{\beta}G_{\theta}(\xi, \xi') = (H_{\theta} g_{\alpha\beta} + \rk_{\theta} \gamma_{\alpha\beta})M \ .  \label{b3}
\end{align}
Then  from   \rf{26},\rf{2431}  we get  for the bosonic and fermionic contributions to the 2-loop  coefficient $f_2$ 
\begin{align} \label{d6}
  \langle \mathcal{L}_{4\rm b} \rangle =&\te  N_{x}\rh_{x}^2-\frac{1}{2}N_{x}^2 \rfe_{x}\rh_{x}-\frac{1}{2}N_{x}(N_{x}+2) \rfe_{x}^2+N_{y}\rh_{y}^2+\frac{1}{2}N_{y}^2 \rfe_{y}\rh_{y} \ ,\\
 \label{d7}
    \langle \mathcal{L}_{4\rm f} \rangle = &
   - \tfrac{1}{2} \rfe_{x} \rh_{\theta} \,N_{x}  {\nt}
    +\tfrac{1}{4}\big[N_x(\rh_x+ \rfe_x )- N_y\rh_y\big]  \rfe_{\theta} {\nt} \nonumber \\
    &\ \ \ + \tfrac{1}{48}\big{[}N_{x}(N_{x}-1)  -N_{y}(N_{y}-1) +  ({\nt}-2)       \big{]}\nt\,   \rh_{\theta} \rfe_{\theta}\nonumber \\
    &\ \ \ -\tfrac{1}{8} \rh_{\theta}^2 \nt^2 - \tfrac{1}{4} \rfe_{\theta}\rk_{\theta} {\nt} \ .
\end{align}
We used that 
\begin{align}
    & \tr(\gamma_{\a}\Gamma_{ij}\Gamma_{123}P_+) = 0 \ ,  \quad  \tr(\gamma_{\a}\Gamma_{a'b'}\Gamma_{123}P_+) = 0 \ , \quad 
     \Gamma_{ij}\Gamma^{ij} = -N_{x}(N_{x}-1) \ ,  \quad  \Gamma_{a'b'}\Gamma^{a'b'} = -N_{y}(N_{y}-1)  , \no\\
    & \langle (\bar{\theta}\gamma_{\alpha}i\mathfrak{D}_{\beta}\theta)(\bar{\theta}\gamma_{\rho}i\mathfrak{D}_{\lambda}\theta)\rangle = \tr(\gamma_{\beta}\gamma_{\alpha}P_+)\tr(\gamma_{\lambda}\gamma_{\rho}P_+) \rh_{\theta}^2-\tr(\gamma_{\lambda}\gamma_{\alpha}\gamma_{\beta}\gamma_{\rho}P_+) \rh_{\theta}^2 \nonumber \\
    &\qquad \qquad \qquad \qquad \qquad \qquad\ -\tr\big[ \gamma_{\alpha}P_+ M\gamma_{\rho}(H_{\theta}g_{\lambda\beta}+\rk_{\theta}\gamma_{\lambda\beta})M\big]  \rfe_{\theta} \ .
\end{align}
Note that the second line in \eqref{d7} vanishes identically for ${N_{x}=3, \, N_{y} =5, \,  {\nt} = 16}$ while the third line is a covariant version of the same  expression in flat space. 


Let us introduce the  following ``equations of motion''  combinations   which 
are  essentially the corresponding  regularized ``$\delta(0)$'' or $\zeta(0)$ values\foot{Note that  in curved 2d space like 
\adst the  $\delta(0)$  terms  contain, in addition to the quadratically divergent part, also a finite part so  that their 2d 
regularized values are non-zero (see e.g. \ci{Randjbar-Daemi:1987rfs}).}
\be
E_x =2 ( \rh_{x}  + \rfe_x) \ , \qquad E_y =2   \rh_y \ , \qquad  E_\theta = 2 \rh_\theta \ ,  \la{b107} 
\ee
and  assume that regularization is such that  in \rf{b3} one has    $\rk_{\theta}=0$.\foot{This should follow  from integrability of Killing spinor derivative.}
Also, let us  set 
\be \la{bb12}
N_{x}(N_{x}-1)  -N_{y}(N_{y}-1)  +( {\nt}-2) =0\ , 
\ee
which is  satisfied  for  the relevant   values  $N_x=3, \ N_y=5, \ \nt=16$. 
Then from   \rf{d6}  and  \rf{d7}  we get 
\begin{align} 
  \langle \mathcal{L}_{4\rm b} + \mathcal{L}_{4\rm f}   \rangle =
  &\te  {1\ov 4}  N_x E_x \big[E_x - (N_x +4) \rfe_x\big]
 + {1\ov 4} N_{y}E_y \big[ E_y +N_{y} \rfe_{y}\big] \no 
 \\ 
 \label{d007}
     &\te  -
     \tfrac{1}{4} \, {\nt}  E_{\theta} \big[ N_{x}   \rfe_{x}  + {1\ov 8} \nt E_\theta \big] 
   +
    \tfrac{1}{8}{\nt} (N_x E_x - N_y E_y) \rfe_{\theta} \ .   
\end{align}
Thus the total 2-loop contribution $ \langle \mathcal{L}_{4\rm b} + \mathcal{L}_{4\rm f}   \rangle $ is proportional 
to the  $E_{x,y,\theta}$  or  ``$\delta(0)$''  terms.

More explicitly, 
for ${N_{x}=3, \, N_{y} =5, \,  {\nt} = 16}$   we get 
\begin{align} 
  \langle \mathcal{L}_{4\rm b} + \mathcal{L}_{4\rm f}   \rangle =& \te
   (- {21\ov 4} E_x   -
   12 E_\t) \rfe_x  + {25\ov 4} E_y \rfe_y  +
    (6 E_x -10 E_y) \rfe_\t\no \\  & + \te 
  {1\ov 4}  ( 3 E_x^2 + 5 E_y^2  -   32 E^2_\t) \ . \la{b108}
  \end{align}
  This, however, may not  represent the full 2-loop  contribution to the free energy: 
  there may be a non-trivial   contribution of  a  local  measure  which is also proportional to 
  $\delta(0)$ terms.\foot{For example, the standard bosonic measure $\prod_\xi \sqrt{G(x(\xi))}$
   would  lead to terms like $\int \delta(0)[  x^2(\xi) + ...]$  contributing 
   $\int \delta(0)\, \langle x^2 \rangle  \sim \delta(0) \rfe_x$  at the 2-loop order
   (see, e.g.,  \ci{Andreev:1990iv}).  The supersymmetric measure  for the GS string 
   may also   contain  terms like $\int \delta(0)[  \bar \theta M  \theta  + ...]$, etc.  
   In general,  the  contribution of the measure may  depend on the induced metric   and thus will not cancel 
   in a ratio of two  Wilson loop expectation values  (like  for circle and  for latitude   which have  different induced metrics).  }
  Here we will not attempt to fix the measure contribution
  (which was not relevant  in the dimensional regularization used in the  main part of the paper). 
      
    The $E^2$  terms  in the second line 
    may  cancel in a 
    {supersymmetric regularization} 
    like the one discussed in \ci{Giombi:2020mhz}  where 
    \be \la{bb33}
     E_\t  = \ha E_x=\ha E_y  \ \ \to \ \   3 E_x^2 + 5 E_y^2 - 32 E^2_\t =0\ . \ee 
        The $\rm G$-terms  in the  first line of \rf{b108}  will   
           contain  log UV divergences; both the   divergent and finite part of the 2-loop 
           result  in \rf{b108} may still be  changed  by the    contribution 
         of the path integral  measure.

     \iffa   
   Let us consider flat space case:   then \rf{d6},\rf{d7} reduces to 
   \begin{align} \label{d67}
  \langle \mathcal{L}_{4\rm b} + \mathcal{L}_{4\rm f}  \rangle =\te  N_{x}\rh_{x}^2  +  N_{y}\rh_{y}^2   -\tfrac{1}{8}  \nt^2  \rh_{\theta}^2
   \ .
\end{align}
In flat space  $N_x + N_y=8$  and $ \rh_{x} = \rh_{y}$  should be just $\delta(0) = {1\ov 4 \pi} L^2$  up to  factor of 2. 
Same should apply to $\delta(0)$ in   Dirac operator. 
Then there is no cancellation unless  coefficient   of $\theta^4$ term is off by  factor of 1/4 or 
the computation  is missing this factor   or definition   of $\rh_\t$ is somehow 2 times  $\delta(0)$ ?! Or we need measure factor? 
  That suggests   that idea     about cancellation of 2nd line in \rf{b108} may be  wrong. 
  Then let us try instead 
   \be \la{bb33}
     E_\t  =  E_x =  E_y  \ee
  \fi


Let us   now  consider the particular case of the spectral 
$\zeta$-function regularization in \adst   (see, e.g., \ci{Camporesi:1991nw,Camporesi:1995fb,Drukker:2000ep,Buchbinder:2014nia,Giombi:2020mhz})\foot{
  Here  $ \zeta(0; \xi,\xi')$ plays the role of the regularized $\delta$-function (the \adst  radius  is set to 1).
  $\zeta_\theta$ corresponds to the squared Dirac operator.}
\begin{align}
&(-\nabla_{\alpha}\nabla^{\alpha}+2)G_{x} =\delta^{(2)} _x( \xi,  \xi')=  \zeta_{m^2=2}(0 \ ;  \xi ,  \xi'), \qquad 
-\nabla_{\alpha}\nabla^{\alpha}G_{y} =\delta^{(2)} _y( \xi,  \xi')  = \zeta_{m^2=0}(0 ;  \xi ,  \xi') \ ,    \\
& \qquad \qquad \qquad i \slashed{\mathfrak{D}}G_{\theta } =\ri   \delta^{(2)} _\theta( \xi,  \xi') = \ri  \zeta_{\theta}(0;  \xi , \xi') \ , 
\\    &\rh_{x}= - \rfe_{x} +  \tfrac{1}{2}\zeta_{m^2=2}(0;  \xi , \xi)  \ ,  \qquad \rh_{y} = \tfrac{1}{2}\zeta_{m^2=0}(0;  \xi,  \xi) 
   \ , \qquad \rh_{\theta} = \tfrac{1}{2}\zeta_{\theta}(0;  \xi ,  \xi), \\
   &   \zeta_{m^2}(0; \xi,\xi) =\tfrac{1}{4\pi} ( \tfrac{1}{6} R^{(2)} - m^2) =    -\tfrac{1}{12\pi } -\tfrac{1}{4\pi}m^2 \ , \\
 & \zeta_{\theta}(0; \xi,\xi) =\tfrac{1}{4\pi} ( \tfrac{1}{6} R^{(2)} -  \tfrac{1}{4} R^{(2)}  - m^2) =  \tfrac{1}{24\pi } -\tfrac{1}{4\pi}m^2   \ \la{b77} . 
\end{align}
Then   we get  the following explicit expressions  for 
the coefficients in \rf{bb1}--\rf{b3} 
\begin{align}
    &\rG_{x} = \tfrac{1}{2\pi}\log \bar \Lambda  -  \tfrac{1}{2\pi}\ , \qquad \rG_y = \tfrac{1}{2\pi}\log \bar \Lambda  \ , \qquad 
     \log \bar \Lambda \equiv \log \Lambda + \Ge \ , \la{b88}\\
    &  \rh_{x} = - \rG_x  -\tfrac{7}{24\pi}  \ , \qquad\ \  \  \rh_{y} = -\tfrac{1}{24\pi} \ ,\la{b99} \\
    &\rG_{\theta} =- \tfrac{1}{2\pi}\log \bar \Lambda  +  \tfrac{1}{4\pi}  
       \ , \qquad  \rh_\theta  = -\tfrac{5}{48\pi} \ , \qquad \rk_{\theta}=0 \ , \qquad H_{\theta} = -\tfrac{5}{48\pi} \ , \la{b66}
\end{align}
where $\Lambda $ is 2d UV cutoff. 
Comparing  this to the dimensional regularization expressions  in \rf{1.8},\rf{9}  we conclude  that $\rG_x$ and $\rG_\theta$ 
 are related by  ${1\ov \ept} \leftrightarrow  2\log \bar \Lambda $
with the finite parts being  the same.

Then for \rf{d6} and \rf{d7} we find 
\begin{align}
 &   \langle \mathcal{L}_{\rm 4b} \rangle = \tfrac{1}{96\pi^2} d_1 \log \bar \Lambda   + \tfrac{1}{576 \pi ^2} d_2 \ , \la{b17} \\
    & d_1 = (7N_{x}^2+28 N_x -N_{y}^2)\Big|_{N_x=3, \, N_y=5} =
      122    \, , \quad 
 d_2= - (42 N_x^2+119 N_x-N_y)\Big|_{N_x=3, \, N_y=5}= - 730\no 
\ , \\
   & \langle \mathcal{L}_{\rm 4f}\rangle =\tfrac{1 }{4608 \pi ^2} e_1 \log \bar \Lambda 
   +\tfrac{1}{18432 \pi ^2}e_2 \ , \la{b18}\\
    & e_1 = \nt  \left(5 N_{x}^2+283 N_{x}-5 N_{y}^2-19 N_{y}+5 \nt-  10\right)\Big|_{N_x=3, \, N_y=5, \, \nt=16} =11904  \ ,\no \\
    &
    e_2 = \nt \left(-10 N_{x}^2-806 N_{x}+10 N_{y}^2+38 N_{y}-35 \nt+20\right)\Big|_{N_x=3, \, N_y=5, \, \nt=16} =-41728\ , \no \\
     &   \langle \mathcal{L}_{\rm 4b}  +  \mathcal{L}_{\rm 4f}\ \rangle = 
     {185\ov 48\pi^2 } \log \bar \Lambda - {113\ov 32\pi^2} \ . \la{b19}
\end{align}
The coefficients of the   divergent term and the finite term are   not the same as in the dimensional regularization in \rf{101},\rf{122}
 which may be attributed to the  still missing contribution of the measure.\foot{In addition, 
  there   may be a  non-trivial   finite contribution of  determinants  of the local ghost operators 
    that  are proportional to ``$\delta(0)$'' terms.} 

\iffa 
going back to  delta(0)^2    story in flat space, one is to
 use  usual delta(0) for fermions but include (1/2)^2  for fermion loop.
(this is of course at the end same as adding 1/2  to propagator in
*this* case).
Then  quartic power divergences cancel.
\fi

\section {1-loop    correction to  2-point function    \la{apC} }

One  could wonder if one is to take into account   possible 1-loop renormalization of the 2-point functions of the fluctuation fields. 
They  are, in fact,     UV finite  in agreement with  the absence of non-trivial 1-loop   divergences in the  GS string  partition function. 

Let us illustrate this  on the example 
of the  1-loop    correction to  the 2-point function   $\langle x^{i}(\w_{1})x^{j}(\w_{2})\rangle$ of the massive  scalar
using dimensional regularization (with $\rd=1+d=2-2\eps$).
\iffa \foot{Alternatively,  we  may read off 
 the coefficients of wave function and mass renormalizations from (\ref{6.8}), but  the resulting 2-loop   contribution (after taking the expectation value of the counterterm) will be the same.}\fi 
\be
\begin{tikzpicture}[thick,baseline=0]
	\begin{feynman}
	\vertex (a1) at (-1,0);
	\vertex (a2) at (1,0);
	\vertex (oo1) at (0,0);
	\vertex (oo2) at (0,1);
	\diagram* {
	(a1) -- (oo1);
	(a2) -- (oo1);
	(oo1) -- [half left] (oo2);
	(oo1) -- [half right] (oo2);
	};
	\end{feynman}
	\node[left] at (a1) {$i$};	
	\node[right] at (a2) {$j$};	
\end{tikzpicture}
\no \ee
We find (see  \rf{2.11},\rf{333}--\rf{335})
\ba
\la{6.8}
&\langle  x^{i}(\w_{1})x^{j}(\w_{2})\rangle_{\rm 1} =  \, \langle x^{i}(\w_{1})x^{j}(\w_{2})\int d^{\rd } \w\,\sqrt{-g}\, \mc L_{4x}\rangle  \\ 
&= - \,\delta^{ij}\int d^{\rd}\w\,\sqrt{-g}\,\Big[
 (N_{x}+4 ) \gxx(\w_1,\w)   \gxx(\w,\w_2)   
 + \tfrac{4+d(4-N_{x})+3N_{x}}{2(d+1)} \del_{\w \a}  \gxx(\w_1,\w)    \del_{\w}^{ \a} \gxx(\w,\w_2)   
\Big]  \, \gxx\ .\no 
\ea
Since $\gxx=\gxx(\w, \w) $ in \rf{4.12}  is constant  we may integrate by parts  and use \rf{4.19} 
to  obtain
\ba
\la{6.10}
\langle & x^{i}(\w_{1})x^{j}(\w_{2})\rangle_{\rm 1} = \delta^{ij} P 
 \int d^\rd\w\,\sqrt{-g}\,  \gxx(\w_1,\w)   \gxx(\w,\w_2)   \ , \ \ \ \  \  P= P_x + ...\ , \ \ \ \  P_x= 2  \, N_x  \frac{1-d}{1+d}\, \gxx\ . 
\ea
The  fermionic loop  contribution  to  this 2-point function  
 contains a  non-zero  contribution   coming from  the $c_1$ term  in \rf{153}.\foot{Here we ignore the terms in \rf{5566} that  give trivial contributions after taking spinor traces. 
 }
 As a result,   the  factor  $P$ in \rf{6.10}  is  changed   to\foot{Note that  $ \langle x^{i}(\w_{1})x^{j}(\w_{2})(\bm{x}^{2}+\tfrac{1}{2}\partial_{\l}\bm{x}\cdot\partial^{\l}\bm{x})\rangle = 0$, \ $ \langle x^{i}(\w_{1})x^{j}(\w_{2})\partial_{\alpha}\bm{x}\cdot\partial_{\beta}\bm{x}\rangle = -\frac{4}{d+1}\eta_{\alpha\beta}\delta^{ij}.$
} 
\be
\la{6.15}
P=\Big(2N_{x}\,\gxx + \frac{1}{1+d}N_{\theta}\, \sss \Big)\frac{1-d}{1+d}\ .
\ee
This  expression is finite in the limit $d\to 1$. 
Taking  $d\to 1$  and setting $N_x=3, \  N_\theta=16$ (and reversing the overall  sign)  we get  for the corresponding finite   1-loop   mass counterterm
\be
\la{6.16}
\delta \mc L_{2\, x} =\ha 
 C_1  \bm{x}^{2}, \qquad \qquad 
 C_1 =  - \lim_{d\to 1} P 
 =  {1\ov 2\pi} \ . 
\ee
Note that  $C_1$  does not  depend on the values of finite parts $a_\rmb$ and $  a_\rmf $  in 
$\gxx$  \rf{1.8} and $\sss$ \rf{9}.\foot{Surprisingly, 
 this value of $C_1$  does not match the   coefficient in the  subleading 
 term in the Bremsstrahlung function
 $B(\lambda)
=B_0 + B_1 + ...= {\sqrt \l \ov 4 \pi^2} - { 3\ov 8 \pi^2}   + ...$ (cf.  \ci{Correa:2012at,   Buchbinder:2013nta,Giombi:2017cqn}).  Since $x^i$ corresponds to the displacement operator
 (with the  2-point function proportional to $12 B$) 
one  would expect  $  C_1$ to be  directly  related  (up to $1\ov 2\pi$    factor)  to  $12 B_1= - {9\ov  2 \pi^2}$.
This may be related to  normalization of the  boundary limit of the 2-point function. We leave the resolution of  this puzzle to the future. }

\iffa 
The contribution of \rf{6.16} to the 2-loop free energy would be  given by its  expectation value  
$
\langle\delta \mc L_{2\, x }\rangle = \ha  N_x C_1 \, \gxx $.  That may be  compared to the coefficients 
of poles   coming from the first and second terms in \rf{5.41}, suggesting the renormalization prescription  in \rf{11}.\foot{The 2-point function 
 correction corresponds to cutting one of the two propagators open, determining the  mass   correction and then subtracting it from the 2-loop contribution according to \rf{11}.}
 Similar discussion applies to the 1-loop   correction to the fermionic propagator   and the corresponding counterterm. 
\fi 


\section{Strong coupling expansion of $1\ov 2 $ BPS  Wilson loop   in ABJM  theory  \la{apD}}

The ABJM theory \cite{Aharony:2008ug} is dual to 
 M-theory in AdS$_{4}\times S^{7}/\ZZ_{k}$ or, in the  't Hooft limit (large $N$ with fixed $\l=N/k$),  to 
 IIA  string on AdS$_{4}\times \CP^{3}$.
 One can try to compare the   localization prediction  for the $\ha$  BPS   circular WL   with its string dual 
given by 
  the disk  partition function of type IIA superstring  in AdS$_{4}\times \CP^{3}$ 
 expanded  near \adst   minimal surface. 
 
 At 1-loop order that was discussed in 
 \ci{Giombi:2020mhz} (see also 
 \ci{Kim:2012tu,Medina-Rincon:2019bcc}).
 A 2-loop  computation  should be possible using  the same methods as in the \adss case in this  paper. In particular, 
since the $S^5$ modes do not appear to play much role  at 2 loops, it is possible 
that the extension to the case of  AdS$_{4}\times \CP^{3}$
may be   straightforward.
\iffa 
\foot{Fermions are a bit different but probably  not too difficult to  do as well.
First trivial check is to see that $1/\eps^2$ cancel  already for bosons.
In both cases $N_y=8-N_{x}$,    and from  3.14 in Stefan's file   one finds
that $\log^2$  div  cancel  for  any $N_{x}$
(that also means for  AdS$_{3}\times S^{3}\times T^{4}$).}
\fi

Here we  will only review the known  expressions for the    strong   coupling expansion
of the WL  expectation value  found  from localization. From the general expression in   \cite{Marino:2009jd,Drukker:2010nc}, 
 taking  the large $N$, fixed $\lambda= N/k$  limit  one finds
 \be\la{811}
\langle W \rangle= \frac{N}{4\pi\l}e^{\pi\sqrt{2(\l- {1\ov 24})}}  +  \mc O(N^{0}) 
 = \frac{N}{4\pi\l}e^{\pi\sqrt{2\l}}    \Big[1-\frac{\pi}{24\sqrt 2}\frac{1}{\sql}+\cdots    \Big]+\mc O(N^{0})\ . 
\ee
Here  the natural  relation  for the dual string tension is\foot{An  indication that this is the correct relation
for the  string tension  comes from the  expressions for the  magnon dispersion relation  and  the cusp anomaly 
at 2-loop order  \cite{Bianchi:2014ada}: the  cusp anomaly   takes a   natural form in terms of the effective tension  given by 
$T= \frac{1}{\sqrt 2}\sqrt{\l-\frac{1}{24}}-\frac{\log 2}{2\pi}$  (the $\log 2$  term is not   relevant  in the WL  case as it  leads just 
to an extra overall  rescaling).
 This is also   suggested by   the structure of localization matrix model that  implies  
   that  the combination $\l- \frac{1}{24} +  \mc O(N^{-1})$     plays  a special role. 
This follows  from the Fermi gas representation   $\exp(-\mu N  + J)$, 
 $J= \frac{1}{3} C \mu^3  + B \mu + A$, \ $B= \frac{k}{24}+\frac{1}{3k}$  (for a review see, e.g.,  \cite{Beccaria:2023ujc}).
This suggests an  effective shift $N\to N-B=  N-\frac{k}{24}+\frac{1}{3k}$. 
 Also,  the argument in \cite{Bergman:2009zh} suggests 
that the AdS  radius in IIA limit should be  given by 
$L^4 = 2 \pi^2 \a'^2 [ \l - \frac{1}{24} (1- {1\ov k^2}) ]$    which at large $N$  for fixed  $\l$ 
  translates into  the  expression for the 
effective string   tension  $T = {1\ov \sqrt 2} \sqrt{\l - {1\ov 24}}$.}
\be
\la{821}
T=  \frac{1}{\sqrt 2}\sqrt{\l- \te {1\ov24}}\ , 
 \qquad\ \ \ \   \ \  T_0\equiv  \frac{\sqrt \l}{\sqrt 2}= T\sqrt{1 + \tfrac{1}{48} T^{-2} } \ . \ \ \ \ 
\ee
Written in terms of $T$  and $g_{\rm s}= \sqrt \pi (2\l)^{5/4} N^{-1} $ as in    \ci{Giombi:2020mhz}    \rf{811} becomes 
\be\la{831}
\langle W \rangle= \frac{\sqrt{T_0} }{\sqrt{2 \pi} \ g_{\rm s} }\ e^{2\pi T }  +  \mc O\big((g_{\rm s})^{0})\big) \ 
= Z_1 \ \sqrt{ T_0\ov T} \, e^{2\pi T }  +  \mc O\big((g_{\rm s})^{0}\big) \  , \qquad \ \  Z_1 = \frac{\sqrt{T} }{\sqrt{2 \pi} \ g_{\rm s} } \ . 
\ee
The corresponding free energy  of the string theory on the disk   should then have the form (cf. \rf{3}) 
\be 
F= - \log \big(Z_1^{-1}  \langle W \rangle \big) = - 2 \pi T   + \ha  \log { T\ov T_0 } =
- 2 \pi T    - \tfrac{1}{  192}  T^{-2}  + \OO(T^{-3}) \ . \la{841}
\ee
 This implies    that the 2-loop  string contribution  here should be zero  -- the first  non-trivial correction to the
  1-loop result should come at 3 loops.

  At fixed $k$ the localization result  expanded in 
  large $N$ or  in the  large M2 brane tension $T_2$ may be written as \ci{Giombi:2023vzu}
  \be\la{85}
\langle W \rangle=   \frac{1}{2\sin \frac{2\pi}{k} }\, e^{ {\pi^2 \ov k} T_2}
\Big[
1-\frac{k^2+32 }{24k }\frac{1}{T_2  }+\OO(T_2^{-2}\big)\Big] \ , \qquad \qquad T_2 = {1\ov \pi} \sqrt{ 2 k N} \ . 
\ee
  Here the prefactor is the 1-loop  correction  to  partition function of M2 brane on AdS$_2\times S^1$  inside of AdS$_{4}\times S^{7}/\ZZ_{k}$
  \ci{Giombi:2023vzu}. 
  The  subleading $1/T_2$ term should be  the 2-loop M2  brane  correction. 
  
To compute the   string  or M2 brane partition function expanded  near  AdS$_2$  or AdS$_2\times S^1 $ 
minimal surface we will need to start with the corresponding GS    or BST   action. 
The type IIA  GS string   action in AdS$_{4}\times \CP^{3}$ expanded   in the static gauge will    have a similar form 
to the \adss  GS action.\foot{The  supercoset construction  of this action was given in \ci{Arutyunov:2008if,Stefanski:2008ik}. It can also be   obtained by  double dimensional reduction of the action for M2 brane in AdS$_{4}\times S^7$  given in 
\ci{deWit:1998yu,Claus:1998fh} (see \ci{Uvarov:2009hf}; this  action in l.c. gauge was used in  2-loop  cusp anomaly computation 
 \cite{Bianchi:2014ada}). 
 A better starting point may be the action in \ci{Gomis:2008jt,Sorokin:2010wn} that has a similar form to the \adss 
   action.
   The quadratic in fermions part of the action was  discussed in \ci{Kim:2012tu}. To get the quartic fermionic term
   one may  use the general form of  expansion in $\theta$  of the  GS action in a IIA  supergravity background 
     given in \ci{Wulff:2013kga}.}

Let us note that there is a recent proposal in \cite{Gautason:2025plx}  suggesting that the  comparison 
 between the gauge theory  localization result and the M2 brane  partition function  should be 
 done differently and as a result  
\iffa 
According to \cite{Gautason:2025plx}    starting with the Fermi  gas expression   for 
$\langle W\rangle$  one should not perform  the integral over $\mu$  variable ``dual'' to $N$   so that 
 $\langle W\rangle = \frac{1}{2\sin\frac{2\pi}{k}}e^{2\mu/k}+\mc O(e^{-\mu})$
where $\mu$    should  be effectively replaced by  $\frac{\pi}{\sqrt 2}\sqrt{Nk}$.
 The exponential factor 
is then $e^{\pi\sqrt{2\l}}$  without a  shift  of  $\l$.
It is thus conjectured that 
\fi 
$\langle W\rangle$  should be   1-loop exact 
in the M2-brane sense, i.e.  there  should  no  M2  brane 2-loop correction  (for fixed $k$).
It would be interesting to  check this by direct 2-loop AdS$_2\times S^1$ M2 brane computation as in the 
recent example  of  M2 brane in  AdS$_7 \times S^4$  expanded  near AdS$_3$ minimal surface  \ci{Beccaria:2025ahf}.

\iffa
In  general, one may prefer to  compute  the full  2-loop expression for the AdS$_2\times S^1 $   M2  brane partition 
in AdS$_{4}\times S^7/Z_k$   with  string theory result been then restriction to the 0-mode sector as at 1-loop level  in \ci{Giombi:2023vzu}. 
Assuming the M2 brane action admits a static $\kappa$-symmetry   gauge in which there are no   3-point  boson-fermion-fermion  couplings 
so that the 2-loop correction is given  again just by OO type diagrams, 
summation over $S^1$ modes   from each loop will factorize   and thus may be straightforward to do. 
\fi


\newpage

\small 
\bibliography{BT-Biblio}
\bibliographystyle{JHEP-v2.9}
\end{document}